\documentclass{article}
\usepackage{ijcai24}

\usepackage{times}
\usepackage{soul}
\usepackage{url}
\usepackage[hidelinks]{hyperref}
\usepackage[utf8]{inputenc}
\usepackage[small]{caption}
\usepackage{graphicx}
\usepackage{amsmath}
\usepackage{amsthm}
\usepackage{booktabs}
\usepackage[switch]{lineno}
\usepackage{dsfont}

\usepackage[ruled,no end, linesnumbered]{algorithm2e}

\usepackage{xspace}
\usepackage[table]{xcolor}
\usepackage{amssymb}
\usepackage{amsfonts}
\usepackage{paralist}
\usepackage[sort&compress,nameinlink,noabbrev]{cleveref}
\usepackage{comment}

\newcommand{\appsymb}{$\star$}
\usepackage{thm-restate}

\usepackage{etoolbox} %

\newcommand{\appendixproofwithstatement}[3]{%
  \gappto{\appendixtext}{
    \subsection{Proof of \cref{#1}}\label{proof:#1}
    #2
    \begin{proof}
    #3\end{proof}
  }
}
\newcommand{\appendixproofwithstatementcontinued}[4]{%
\begin{proof}[Proof Sketch]
#3
\end{proof}
  \gappto{\appendixtext}{
    \subsection{Continuation of Proof of \cref{#1}}\label{proof:#1}
    #2
    \begin{proof}[Proof (Continued)]
    #4\end{proof}
  }
}

\newcommand{\appendixsection}[1]{%
  \gappto{\appendixtext}{
    \section{Additional material for Section~\ref{#1}}
    \label{appsec:#1}
  }
}

\usepackage{graphicx}

\usepackage[textsize=tiny,textwidth=1.5cm,linecolor=green!70!black, backgroundcolor=green!10, bordercolor=black,disable=true]{todonotes}%
\usepackage{hyperref} 
\hypersetup{%
 backref=true, 
 pagebackref=true, 
 hypertexnames=true,
 colorlinks=true,citecolor=green!35!black,linkcolor=red!60!black,
 citecolor = purple%
}

\usepackage{pgfkeys}

\usepackage{mdframed}

\newcommand{\probname}[1]{\textsc{#1}}

\newcommand{\wrt}{with respect to\xspace}
\newcommand{\ifDirection}{\textbf{(If).}\xspace}
\newcommand{\onlyifDirection}{\textbf{(Only if).}\xspace}

\newcommand{\phaseVar}{b}
\newcommand{\startVar}{\mathsf{start}}
\newcommand{\ndVar}{\mathsf{end}}
\newcommand{\thetaTwoP}{\ensuremath{\Theta_2^{\text{P}}}}
\newcommand{\thetaTwoPComplete}{\thetaTwoP\text{-\normalfont{complete}}\xspace}

\newcommand{\thetaTwoPHardness}{\thetaTwoP\text{-\normalfont{hardness}}\xspace}
\newcommand{\NPhard}{\NP\text{-\normalfont{hard}}\xspace}
\newcommand{\DPhard}{\DP\text{-\normalfont{hard}}\xspace}
\newcommand{\NPcomplete}{\NP\text{-\normalfont{complete}}\xspace}

\newcommand{\NP}{\text{\normalfont{NP}}}
\newcommand{\DP}{\text{\normalfont{DP}}}
\newcommand{\coNP}{\text{\normalfont{coNP}}}
\newcommand{\PP}{\text{\normalfont{P}}}

\newcommand{\FPT}{\text{\normalfont{FPT}}\xspace}
\newcommand{\XP}{\text{\normalfont{XP}}\xspace}

\newcommand{\sigmaC}{\ensuremath{\Sigma_2^{\text{P}}}}

\newcommand{\MOSP}{MOS\xspace}

\newcommand{\objMOSPDec}[1]{\probname{#1-\MOSP-Dec}}
\newcommand{\objMOSP}[1]{\probname{#1-\MOSP}}

\newcommand{\makespanMOSP}{\objMOSP{\makespanObj}}
\newcommand{\sumCMOSP}{\objMOSP{\sumCObj}}

\newcommand{\makespanMOSPDec}{\objMOSPDec{\makespanObj}}
\newcommand{\sumCMOSPDec}{\objMOSPDec{\sumCObj}}

\newcommand{\UnaryBinPacking}{\probname{Unary-BinPacking}}
\newcommand{\MOSPLS}{\probname{\makespanMOSP-LocalSchedules}}
\newcommand{\MOSPLSshort}{\probname{\makespanMOSP-LS}}
\newcommand{\minMakespan}{\probname{Min\makespanObj}}

\newcommand{\makespanObj}{\ensuremath{\completionTime_{\mathsf{max}}}\xspace}
\newcommand{\sumCObj}{\ensuremath{\completionTime_{\Sigma}}\xspace}
\newcommand{\sumCshort}{$\Sigma$-time}
\newcommand{\optimallocalmakespan}{optimal local-makespan}
\newcommand{\optimallocalsumC}{optimal local-\sumCshort}

\newcommand{\loc}{\mathsf{lo}}
\newcommand{\glob}{\mathsf{g}}

\newcommand{\obj}{\ensuremath{\Omega}\xspace}
\newcommand{\inst}{\ensuremath{I}\xspace}

\newcommand{\nbOrg}{\ensuremath{k}\xspace}
\newcommand{\setOrg}{\ensuremath{\mathcal{O}}\xspace}
\newcommand{\org}{\ensuremath{O}\xspace}

\newcommand{\setTask}{\ensuremath{\mathcal{J}}\xspace}
\newcommand{\setTaskOrg}[1]{\ensuremath{J_{#1}}\xspace}
\newcommand{\nbTask}{\ensuremath{n}\xspace}
\newcommand{\nbTaskOrg}[1]{\ensuremath{\nbTask_{#1}}\xspace}
\newcommand{\task}{\ensuremath{\alpha}\xspace}
\newcommand{\taskOrg}[2]{\ensuremath{\task^{#1}_{#2}}\xspace}
\newcommand{\maxNbTask}{\ensuremath{\nbTask_{\mathsf{max}}}\xspace}
\newcommand{\maxNbMachine}{\ensuremath{\nbMachine_{\mathsf{max}}}\xspace}

\newcommand{\procTime}{\ensuremath{\mathsf{p}}}
\newcommand{\procTimeTask}[2]{\ensuremath{\procTime^{#1}_{#2}}\xspace}
\newcommand{\maxProcTime}{\ensuremath{\procTime_{\mathsf{max}}}\xspace}

\newcommand{\nbMachine}{\ensuremath{\mathsf{m}}\xspace}
\newcommand{\setMachine}{\ensuremath{\mathcal{M}}\xspace}
\newcommand{\setMachineOrg}[1]{\ensuremath{M_{#1}}\xspace}
\newcommand{\nbMachineOrg}[1]{\ensuremath{\nbMachine_{#1}}\xspace}

\newcommand{\targetValue}{\ensuremath{\tau}\xspace}

\newcommand{\completionTime}{\ensuremath{\mathsf{C}}}

\newcommand{\completionTimeTaskSchedule}[3]{\ensuremath{\completionTime^{#1}_{#2}(#3)}\xspace}
\newcommand{\machineTaskSchedule}[3]{\ensuremath{\nbMachine^{#1}_{#2}(#3)}\xspace}
\newcommand{\makespanSched}[1]{\ensuremath{\completionTime_{\mathsf{max}}(#1)}\xspace}
\newcommand{\sumCSched}[1]{\ensuremath{\completionTime_{\Sigma}(#1)}\xspace}
\newcommand{\localMakespan}[1]{\optLocalMakespan{#1}}

\newcommand{\makespanOrgSched}[2]{\ensuremath{\completionTime_{\mathsf{max}}^{#1}(#2)}\xspace}
\newcommand{\sumCOrgSched}[2]{\ensuremath{\completionTime_{\Sigma}^{#1}(#2)}\xspace}

\newcommand{\optLocalMakespan}[1]{\ensuremath{\mathsf{OPT}\text{-}\completionTime_{\mathsf{max}}^{#1}}\xspace}
\newcommand{\optLocalSumC}[1]{\ensuremath{\mathsf{OPT}\text{-}\completionTime_{\Sigma}^{#1}}\xspace}

\newcommand{\sched}{\ensuremath{\sigma}\xspace}

\newcommand{\triplet}{\ensuremath{T}\xspace}
\newcommand{\sumTriplet}{\ensuremath{B}\xspace}
\newcommand{\setInt}{\ensuremath{X}\xspace}
\newcommand{\nbTriplet}{\ensuremath{q}\xspace}
\newcommand{\nbTripletMax}[1]{\ensuremath{\nbTriplet^{\max}_{#1}}\xspace}
\newcommand{\totsum}{\ensuremath{Q}\xspace}

\newcommand{\instTP}{\ensuremath{\mathcal{I}}\xspace}
\newcommand{\instTPInd}[1]{\ensuremath{\instTP_{#1}}\xspace}
\newcommand{\instTPPrime}{\ensuremath{\mathcal{I}'}\xspace}
\newcommand{\instTPPrimeInd}[1]{\ensuremath{\instTPPrime_{#1}}\xspace}

\newcommand{\superscriptTimes}[1]{\ensuremath{^{\otimes #1}}}

\newcommand{\instTPTimes}[2]{\ensuremath{(#1)\superscriptTimes{#2}}\xspace}

\newcommand{\setInstOne}{\ensuremath{\mathcal{S}^{\instTP}}\xspace}
\newcommand{\setInstPrime}{\ensuremath{\mathcal{S}^{\instTPPrime}}\xspace}

\newcommand{\nbInst}{\ensuremath{a}\xspace}
\newcommand{\nbInstPrime}{\ensuremath{a'}\xspace}

\newcommand{\answer}[1]{\ensuremath{\chi(#1)}\xspace}
\newcommand{\integer}{\ensuremath{x}\xspace}

\newcommand{\ThreePartition}{\probname{3-Partition}\xspace}
\newcommand{\ThreePartitionComparison}{\probname{3-Partition Comparison}\xspace}

\newcommand{\nbTripletInst}[1]{\ensuremath{\nbTriplet(#1)}\xspace}
\newcommand{\sumTripletInst}[1]{\ensuremath{\sumTriplet(#1)}\xspace}
\newcommand{\sumTripletInd}[1]{\ensuremath{\mathcal{\sumTriplet}_{#1}}\xspace}
\newcommand{\integerInst}[2]{\ensuremath{\integer_{#2}(#1)}\xspace}

\newcommand{\setIntegerOne}[1]{\ensuremath{\mathcal{X}_{1}^{#1}\xspace}}
\newcommand{\setIntegerTwo}[1]{\ensuremath{\mathcal{X}_{2}^{#1}\xspace}}
\newcommand{\setEqualizingThree}[1]{\ensuremath{\mathcal{E}_{3}^{#1}\xspace}}

\newcommand{\setBin}{\ensuremath{S}\xspace}
\newcommand{\nbBin}{\ensuremath{m}\xspace}

\newcommand{\lengthOffset}[1]{\ensuremath{C(#1)}\xspace}
\newcommand{\ndP}{\mathsf{end}}
\newcommand{\phase}{\mathsf{phase}}

\newcommand{\UBOrgTwo}[1]{\ensuremath{k(#1)}\xspace}

\newcommand{\decprob}[3]{
  \begin{center}%
    \begin{minipage}{0.94\linewidth}%
      \textsc{#1}\\[0.2ex]
      \textbf{Input:} #2\\[0.2ex]
      \textbf{Question:} #3
    \end{minipage}%
  \end{center}
}

\newcommand{\taskprob}[3]{
  \begin{center}%
    \begin{minipage}{0.94\linewidth}%
      \textsc{#1}\\[0.2ex]
      \textbf{Input:} #2\\[0.2ex]
      \textbf{Task:} #3\\
    \end{minipage}%
  \end{center}
}

\newcommand{\myemph}[1]{{\color{green!30!black}\emph{#1}}}
\newcommand{\mypara}[1]{
  \smallskip
  \noindent {\textbf{#1}}}

\newcommand{\intuition}[1]{\textit{#1}}

\newtheorem{definition}{Definition}

\newtheorem{theorem}{Theorem}
\newtheorem{obs}{Observation}
\newtheorem{example}{Example}
\newtheorem{proposition}{Proposition}

\newtheorem{claim}{Claim}[theorem]

\crefname{section}{Section}{Sections}
\crefname{figure}{Figure}{Figures}
\crefname{lemma}{Lemma}{Lemmas}
\crefname{proposition}{Proposition}{Propositions}
\crefname{algorithm}{Algorithm}{Algorithms}
\crefname{obs}{Observation}{Observations}
\crefname{table}{Table}{Tables}
\crefname{theorem}{Theorem}{Theorems}
\crefname{claim}{Claim}{Claims}
\crefname{example}{Example}{Examples}
\crefname{exampleC}{Example}{Examples}
\crefname{corollary}{Corollary}{Corollaries}

\DeclareMathOperator*{\argmin}{arg\,min}

\usetikzlibrary{decorations.pathreplacing,calligraphy,bending}

\newcommand{\taskColorTikz}[5]{
  \filldraw[rounded corners,line width=0.5pt,#5] (#3-#2,0.1) rectangle  (#3,0.65);
  \draw (#3-#2/2,0.35) node {{#1}};
}
      
\tikzset{colorOne/.style={color=blue!60!black, fill=blue!20, dashed}}
\tikzset{colorTwo/.style={color=red!60!black, fill=red!20}}
\tikzset{colorThree/.style={color=teal!60!black, fill=teal!20}}
\tikzset{colorFour/.style={color=brown!60!black, fill=brown!20}}
\tikzset{colorFive/.style={color=violet!60!black, fill=violet!20}}
\tikzset{colorEqualizing/.style={color=black, fill=black!20}}

\newcommand{\mytitle}{Multi-Organizational Scheduling:\\
  Individual Rationality, Optimality, and Complexity}
\title{\mytitle}
\newcommand{\appendixtitle}{Supplementary Material for the Paper ``\mytitle''}

\author{
Jiehua Chen$^1$
\and
Martin Durand$^{1,2}$\and
Christian Hatschka$^1$\\
\affiliations
$^1$TU Wien, Institute of Logic and Computation, Vienna, Austria\\
$^2$Sorbonne Universit\'{e}, CNRS, LIP6, Paris, France
\emails
\{jchen, chatschka\}@ac.tuwien.ac.at,
martin.durand@lip6.fr
}

\begin{document}

\maketitle
\begin{abstract}
  We investigate multi-organizational scheduling problems,
  building upon the framework introduced by~\citeauthor{pascual2009introduction}~\shortcite{pascual2009introduction}.
  In this setting, multiple organizations each own a set of identical machines and sequential jobs with distinct processing times.
  The challenge lies in optimally assigning jobs across organizations' machines to minimize the overall makespan while ensuring no organization's performance deteriorates.
  To formalize this fairness constraint, we introduce \myemph{individual rationality},
  a game-theoretic concept that guarantees each organization benefits from participation.

  Our analysis reveals that finding an individually rational schedule with minimum makespan is $\thetaTwoP$-hard, placing it in a complexity class strictly harder than both NP and coNP.
  We further extend the model by considering an alternative objective: minimizing the sum of job completion times, both within individual organizations and across the entire system.
  The corresponding decision variant proves to be NP-complete.
  Through comprehensive parameterized complexity analysis of both problems,
  we provide new insights into these computationally challenging multi-organizational scheduling scenarios.
\end{abstract}

\section{Introduction}
\label{sec:intro}

Multi-organizational scheduling~(\MOSP) has emerged as a crucial paradigm in distributed computing environments, where organizations collaborate by sharing their computational resources to optimize job processing~\cite{pascual2009introduction}.
In this model, multiple organizations, each possessing their own machines and jobs, connect their resources through a grid network to create more efficient scheduling solutions.
Collaboration can potentially improve global performance metrics, such as the overall \myemph{makespan} (i.e., completion time of the last job) or the \myemph{total sum of completion times of all jobs}.
A possible real-life application of such a model would be a set of universities or research units, each owning a cluster of machines used by their respective members to run computer programs. These organizations may be willing to mutualize their resources in order to balance the computational load of their clusters.

However, this collaborative framework introduces strategic considerations, as each organization prioritizes its own objectives (e.g., minimizing its local makespan).
From a game-theoretical perspective, if any organization would achieve worse performance in the collaborative schedule compared to operating independently,
it is unfair for that organization; so the organization has an incentive to withdraw from the cooperation.
Such a withdrawal could cascade into disrupting other organizations' schedules.
Therefore, a fundamental constraint in our scheduling problem is \myemph{individual rationality}, ensuring that no organization performs worse under cooperation than it would independently.

While individually rational schedules are guaranteed to exist (as organizations can always default to an optimal local schedule),
the challenge lies in finding one that additionally optimizes global performance metrics.
This work focuses on two fundamental metrics: the maximum completion time ($\makespanObj$) and the sum of completion times ($\sumCObj$).
We examine scenarios where both individual organizations and the grand coalition as a whole optimize either $\makespanObj$ or $\sumCObj$, leading to two distinct optimization problems: \makespanMOSP\ and \sumCMOSP\ (formally defined in \cref{sec:prelim}).

\paragraph{Main contributions.}
We introduce \myemph{individual rationality} to the multi-organizational scheduling framework. %
Under this fairness concept, no organization has an incentive to withdraw from collaboration
since no local schedule can achieve a better performance.

We systematically investigate the algorithmic complexity of two optimization problems:  
\makespanMOSP\ and \sumCMOSP.
Generally speaking, both problems are computationally hard.
More precisely, the decision variant of \makespanMOSP\ is $\thetaTwoPComplete$\footnote{\myemph{$\thetaTwoP$} (aka.\ $\PP^{\NP[\log]}$ and $\PP^{\NP}_{\mid\mid}$) is a complexity class, consisting of all problems which can be decided
in polynomial time with \emph{logarithmically} many queries to an NP-oracle~\cite{WagnerBoundeQueries1990}, positioning it between $\NP$ and $\sigmaC$ in the complexity hierarchy.} %
while the one of \sumCMOSP\ is NP-complete.

We also present parameterized complexity analysis considering key parameters (and their combinations). %
They are:
\begin{compactitem}[--]
  \item \myemph{$\nbOrg$:} the number of organizations,
  \item \myemph{$\nbMachine$}: the number of machines,
  \item \myemph{$\nbTask$}: the number of jobs,
  \item \myemph{$\targetValue$}: the target value of the objective (i.e., either makespan or sum of competition times) function in the decision variant,
  \item \myemph{$\maxProcTime$}: the maximum processing time of a job,
  \item \myemph{$\maxNbTask$}: the maximum number of jobs owned by an organization, and 
  \item \myemph{$\maxNbMachine$}: the maximum number of machines owned by an organization.
\end{compactitem}
Note that we chose parameters that are studied in the literature on scheduling~\cite{Mnich15,mnich2018parameterized}, as well as parameters that are unique to the multi-organizational setting:
$\nbOrg,\maxNbTask$, and $\maxNbMachine$. The latter two are localized versions of the global parameters $\nbTask$ and $\nbMachine$, respectively.%

Among the parameterized findings,
for the parameter combination~$\nbOrg+\maxProcTime$, we develop a \myemph{fixed-parameter tractable} (\FPT) algorithm for \makespanMOSP, based on integer-linear programming (ILP).
Our approach is based on an FPT-algorithm by~\citeauthor{Mnich15}~\shortcite{Mnich15} for the classical problem of minimizing the makespan; this is equivalent to our model with a single organization. 
They proved the existence of an optimal solution that evenly distributes all jobs of the same processing time among the machines; the difference is upper-bounded by a function in~$\maxProcTime$ only.
This implies that there are only a few number of different types of machines.
We extend this idea and show that the number of different machine types is upper-bounded by $\nbOrg+\maxProcTime$.
We can then introduce an integer variable for each machine type and use ILP to find an optimal solution.

Via a straightforward dynamic programming (DP) approach, we also demonstrate that \makespanMOSP\ (resp. \sumCMOSP) is in $\XP$
\wrt\ $\nbMachine$ (resp.\ $\nbMachine+\maxProcTime$).
For the hardness, we prove that \makespanMOSP\ remains \DP-hard even when $\nbOrg$ is a small constant\footnote{\DP\ is the class of problems expressible as the difference of an \NP- and a \coNP\ problem~\cite[Chapter 17]{papadimitrioubook}.}.

\cref{tab:result_table} summarizes our complete complexity findings. Due to space constraints, proofs of results marked with a (\appsymb) symbol are deferred to the appendix.

\mypara{Related work.}
\citeauthor{pascual2009introduction}\shortcite{pascual2009introduction}
initiated the study of cooperation in multi-organizational scheduling, where jobs may require parallel execution across machines.
They addressed the problem of minimizing the global makespan under a \myemph{local constraint} that no organization performs worse compared to a specific local schedule, computed using a heuristic.
However, this constraint differs from the individual rationality we focus on in this paper, as the heuristic-based local schedule may not be optimal, meaning organizations might still have an incentive to leave the cooperation.
\citeauthor{pascual2009introduction}~\shortcite{pascual2009introduction} showed that their problem is NP-hard and provided approximation algorithms.

\citeauthor{cohen2011multi}~\shortcite{cohen2011multi} considered the same model and proposed approximation algorithms for sequential jobs.
\citeauthor{durand2021efficiency}~\shortcite{durand2021efficiency} examined a more general setting where the local schedules are given as input and studied its approximability.
Variants of these problems also allow organizations to pursue objectives beyond minimizing the makespan of their own jobs, such as minimizing the sum of job completion times~\cite{cohen2011multi} or the energy required to schedule jobs~\cite{cohen2014energy}.
Other studies relaxed the individual rationality constraint, allowing organizations to accept schedules where their makespan increases, provided the increase is within a given factor~\cite{ooshita2009generalized,ooshita2012price,chakravorty2013algorithms,cordeiro2011tight}.
\citeauthor{rzadca2007scheduling}~\shortcite{rzadca2007scheduling} introduces the notion of self-reliance and 
\citeauthor{skowron2014fair}~\shortcite{skowron2014fair} employed cooperative game theory in multi-organizational scheduling, but using Shapley values as a measure of fairness. As mentioned earlier, our definition of individual rationality is stronger as it compares each organization's outcome to its optimal local schedule, in line with standard individual rationality definitions in coalition formation games.

Parameterized complexity has recently gained attention in scheduling~\cite{mnich2018parameterized}.
In the classical setting (with a single organization), the problem of minimizing the makespan is shown to be FPT \wrt~the maximum processing time of a task $\maxProcTime$~\cite{Mnich15,knop2020combinatorial}. 
Multi-organizational cooperation has also been studied in other contexts, such as matching~\cite{Biro19,Gourves12} and kidney exchange~\cite{SonmezKidney,AR12Kidney,Ashlagikidney,Klimentova21mKEP}

\section{Preliminaries}
\label{sec:prelim}
For details and definitions from parameterized complexity, we refer to the textbook by \citeauthor{CyFoKoLoMaPiPiSa2015}~\shortcite{CyFoKoLoMaPiPiSa2015}. 

Given an integer~$z\in \mathds{Z}$, let $[z]=\{1,\dots,z\}$.
An instance of \MOSP{} is
a tuple~\myemph{$\langle\setOrg, (\setMachineOrg{i})_{i\in [\nbOrg]},
(\setTaskOrg{i})_{i\in [\nbOrg]},
(\procTimeTask{i}{j})_{i\in [\nbOrg], j\in [\nbTaskOrg{i}]}\rangle$},
where \myemph{$\setOrg$} denotes a set of organizations with \myemph{$|\setOrg|=\nbOrg$} such that
for each~$i\in [\nbOrg]$,  
\begin{compactitem}[--]
  \item \myemph{$\setMachineOrg{i}$} denotes a non-empty set of \myemph{$\nbMachineOrg{i}$}$=\!|\setMachineOrg{i}|$ identical machines,
  \item \myemph{$\setTaskOrg{i}$} denotes of a set of \myemph{$\nbTaskOrg{i}$} $=|\setTaskOrg{i}|$ non-preemptive (i.e., each job has to be completely processed before another job) and sequential jobs~$\taskOrg{i}{j}$, and
  \item for each {$j\in [\nbTaskOrg{i}]$}, \myemph{$\procTimeTask{i}{j}$} denotes the processing time of the $j^{\text{th}}$ job\footnote{In this paper we suppose that the instance is encoded in unary.}, called~\myemph{$\taskOrg{i}{j}$} in~$\setTaskOrg{i}$, 
\end{compactitem}
all associated with organization~$\org_i$.

Throughout, we assume that \myemph{$\inst$} denotes an instance of \MOSP{} of the form~$\langle\setOrg, (\setMachineOrg{i})_{i\in [\nbOrg]},
(\setTaskOrg{i})_{i\in [\nbOrg]},
(\procTimeTask{i}{j})_{i\in [\nbOrg], j\in [\nbTaskOrg{i}]}\rangle$.
Moreover, 
we denote
by \myemph{\setMachine} the set of all machines,
i.e., $\setMachine=\bigcup_{i \in [\nbOrg]} \setMachineOrg{i}$,
by \myemph{$\setTask$} the set of all jobs, i.e., $\setTask=\bigcup_{i \in [\nbOrg]} \setTaskOrg{i}$,
by \myemph{$\nbTask$} the total number of jobs,
and finally by \myemph{$\nbMachine$} the total number of machines.

\mypara{Feasible schedules.} A \myemph{schedule}~\myemph{$\sched$} $\colon \setTask \to \setMachine \times \mathds{N}$ is a function that assigns to each job a machine and a completion time.
For notational convenience, for each~organization~$\org_i$ and each~$j \in [\nbTaskOrg{i}]$, 
we denote
by  \myemph{$\machineTaskSchedule{i}{j}{\sched}$} the scheduled machine 
and 
by \myemph{$\completionTimeTaskSchedule{i}{j}{\sched}$} the scheduled completion time of the $j^{\text{th}}$ job of organization~$i$ under~$\sched$.

A schedule is \myemph{feasible} if each job is assigned a machine with feasible completion time and no two jobs can occupy the same machine in the same processing time.
Formally, 
we have that
\begin{compactitem}[--]
  \item for each job~$\taskOrg{i}{j}$, $\completionTimeTaskSchedule{i}{j}{\sched} \ge \procTimeTask{i}{j}$
and %
\item for each two jobs~$\taskOrg{i}{j}\neq \taskOrg{i'}{j'}$ scheduled on the same machine,
$|\completionTimeTaskSchedule{i}{j}{\sched}- \completionTimeTaskSchedule{i'}{j'}{\sched}| \ge \min(\procTimeTask{i}{j}, \procTimeTask{i'}{j'})$, %
i.e., either the starting time of $\taskOrg{i}{j}$ is later or equal than the completion time of $\taskOrg{i'}{j'}$ or the completion time of $\taskOrg{i}{j}$ is earlier or equal to the starting time of $\taskOrg{i'}{j'}$.
\end{compactitem}
A schedule~$\sched$ is a \myemph{local} schedule for \myemph{organization~$\org_i\in \setOrg$} if it is feasible and for every job~$\taskOrg{i}{j}\in \setTaskOrg{i}$ it holds that $\machineTaskSchedule{i}{j}{\sched} \in \setMachineOrg{i}$. 
We will oftentimes just use ``schedules'' to refer to ``feasible schedules'' when it is clear from the context. 

\mypara{Makespan and sum of completion times.}
Let~$\sched$ be a schedule. 
Then, the \myemph{makespan} (resp.\ the \myemph{sum of completion times}, in short \myemph{\sumCshort}) of a set of jobs~\myemph{$\setTask'$} \wrt $\sched$,
denoted as \myemph{$\makespanSched{\sched,\setTask'}$} (resp.\  \myemph{$\sumCSched{\sched,\setTask'}$}), is the maximum completion time~(resp.\ sum of completion times) of all jobs in~$\setTask'$.
The \myemph{makespan} (resp.\ the \myemph{\sumCshort}) of~\myemph{organization~$\org_i\in \setOrg$} \wrt $\sched$ is
\myemph{$\makespanOrgSched{i}{\sched}$} $=\makespanSched{\sched,\setTaskOrg{i}}$ (resp.\  \myemph{$\sumCOrgSched{i}{\sched}$} $=\sumCSched{\sched,\setTaskOrg{i}}$).
We omit the second argument~$\setTask'$ when we refer to the makespan (resp.\ \sumCshort) of all jobs, i.e.,
$\myemph{\makespanObj(\sched)} = \makespanSched{\sched, \setTask}$
and $\myemph{\sumCObj(\sched)} = \sumCSched{\sched, \setTask}$, respectively.

The \myemph{\optimallocalmakespan} (resp.\ \myemph{\optimallocalsumC}) of \myemph{organization~$\org_i$},
denoted as \myemph{$\optLocalMakespan{i}$} (resp.\ \myemph{$\optLocalSumC{i}$})
is the minimum over the makespans (resp.\ minimum over the \sumCshort{s}) of all \emph{local} schedules of~$\org_i$.
In other words, the \optimallocalmakespan\ (resp.\ \optimallocalsumC) of an organization~$\org_i$ is the minimum makespan (resp. minimum \sumCshort) achievable by any schedule where all jobs of~$\org_i$ are only scheduled to the machines of~$\org_i$.

A schedule~$\sched$ is an \myemph{optimal local schedule} for organization~$\org_i$ if it is a local schedule for~$\org_i$ and has makespan equal to~$\optLocalMakespan{i}$ (resp.\ \sumCshort\ equal to~$\optLocalSumC{i}$). 

\begin{example}
\label{ex:1_basic}
We consider an example with two organizations: $\org_1$ and $\org_2$. Organization $\org_1$ owns $\nbMachine_1=2$ machines and $\nbTask_1=3$ jobs with processing times $\procTimeTask{1}{1}=\procTimeTask{1}{2}=\procTimeTask{1}{3}=3$. Organization $\org_2$ owns $\nbMachine_2=1$ machine and $\nbTask_2=6$ jobs with processing times $\procTimeTask{2}{1}=\procTimeTask{2}{2}=\dots=\procTimeTask{2}{6}=1$. Possible local schedules for the instance are drawn in~\cref{fig:ex_1}. 
\begin{figure}[t!] 
\centering
\begin{small}

\begin{tikzpicture}[scale=0.85]
  \begin{scope}[yshift=1.2cm]
    \taskColorTikz{\taskOrg{1}{1}}{3}{3}{1.5}{colorOne}
    \taskColorTikz{\taskOrg{1}{3}}{3}{6}{1.5}{colorOne}
    \begin{scope}[yshift=-0.55cm]
      \taskColorTikz{\taskOrg{1}{2}}{3}{3}{1}{colorOne}
      \node at (-0.75,0.65) {$m_1=2$};
    \end{scope}
  \end{scope}
  
  \begin{scope}[yshift=0cm]
    \taskColorTikz{\taskOrg{2}{1}}{1}{1}{0}{colorTwo}
    \taskColorTikz{\taskOrg{2}{2}}{1}{2}{0}{colorTwo}
    \taskColorTikz{\taskOrg{2}{3}}{1}{3}{0}{colorTwo}
    \taskColorTikz{\taskOrg{2}{4}}{1}{4}{0}{colorTwo}
    \taskColorTikz{\taskOrg{2}{5}}{1}{5}{0}{colorTwo}
    \taskColorTikz{\taskOrg{2}{6}}{1}{6}{0}{colorTwo}

    \node at (-0.75,0.35) {$m_2=1$};
  \end{scope}

  \draw[->] (0,0) -- (6+0.2,0);
  \draw (6,0.1) -- (6,-0.1) node[below]{$6$};
  \draw (0,0.1) -- (0,-0.1) node[below]{$0$};		
\end{tikzpicture}
\end{small}

\caption{Possible local schedules for $\org_1$ (top) and $\org_2$ (bottom) from \cref{ex:1_basic}. Interpretation: Each job is represented by a rectangle, with the length depicting the processing time.
  Jobs on the same row are assigned to the same machine.
  Time goes from left to right, i.e., a job represented left to another job is processed earlier in the schedule.
  $\org_1$'s jobs are in blue, while $\org_2$'s jobs in red.}
\label{fig:ex_1}
\end{figure}
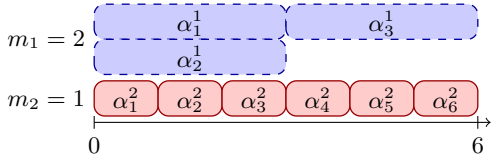
\end{example}

\mypara{Individual rationality.}
A schedule~$\sched$ is called \myemph{individually rational} if
no organization is worse-off by looking at the optimal local-makespan or local-sum of completion times.
More precisely, for the objective of minimizing the makespan ($\makespanObj$),
we require that
$\makespanOrgSched{i}{\sched} \le \optLocalMakespan{i}$ holds for each~$\org_i\in \setOrg$,
while for the objective of minimizing the sum of competition times ($\sumCObj$),
we require that
$\sumCOrgSched{i}{\sched} \le \optLocalSumC{i}$ holds for each~$\org_i\in \setOrg$,

Clearly, for both objectives, individually rational schedules exist as one can compute an optimal schedule for each organization and combine them into a global one. 

\begin{example}
  The local schedules displayed in~\cref{fig:ex_1} are optimal local schedules for both organizations and both objectives.
  The optimal local-makespans of both organizations are $\optLocalMakespan{1}=\optLocalMakespan{2}=6$.
  By definition, this is an individually rational schedule.
  We consider a schedule \sched in which two jobs of~$\org_2$ are first on each machine followed by one of the three jobs of~$\org_1$,
  then \sched  is individually rational and minimizes the overall makespan:
  The processing times of all jobs sum up to~$15$ and we have $3$ machines.
  So the minimum makespan is at least $5$. Moreover, the makespan of~$\org_1$ is $5$ while the makespan of~$\org_2$ is $2$.
  
  One can check that \sched is also optimal when we aim at minimizing \sumCshort\ instead, with a total of $24=3+6+15$. However it is not individually rational if \sumCshort\ is the objective. Indeed, the \optimallocalsumC\ of the organizations are $\optLocalSumC{1}=3+3+6=12$ and $\optLocalSumC{2}=1+2+3+4+5+6=21$ and the sum of completion times of $\org_1$ in \sched is 15. %
\end{example}

\mypara{Central problems.}
We look at two optimization problems, which aim for an optimal solution among all individually rational schedules.
In the following, let $\obj\in \{\makespanObj, \sumCObj\}$. 

\taskprob{\objMOSP{$\obj$}}
{An instance $\inst$ %
  of \MOSP.} %
{Find a schedule~$\sched$ among all \emph{individually rational} schedules for~$\inst$ 
  such that $\obj(\sched)$ is minimum.}

\noindent The decision variants, called \objMOSPDec{$\obj$} have additionally a non-negative integer~$\targetValue$ as input and ask whether there exists an \emph{individually rational} schedule~$\sched$ with $\obj(\sched)\le \targetValue$.

\mypara{Remarks.} Note that in the classical setting (i.e., when the number of organizations is one),
it is NP-hard to find a schedule with minimum makespan
whereas for  the minimum \sumCshort\ case it is polynomial-time solvable~\cite{brucker1999scheduling}.
Hence, \makespanMOSPDec\ is contained in $\Sigma^{\text{P}}_2$ while \sumCMOSPDec\ in NP.
We will show that \makespanMOSPDec\ is between NP and $\Sigma^{\text{P}}_2$; it is $\thetaTwoP$-complete (\cref{thm:makespan_ThetaTwoP}),
while \sumCMOSPDec\ is NP-complete (\cref{thm:avg_NP}).

\newcommand{\cohenfollows}{$^{\dagger}$}
\newcommand{\jansenfollows}{$^{\spadesuit}$}

\begin{table}
\centering
{
  {
    \renewcommand{\arraystretch}{1.2}
    \begin{tabular}{@{}r| r@{\;}c @{\;\;}c@{\;\;} r@{\;}c@{}}
      \toprule
      & \multicolumn{2}{c}{\makespanMOSP}
      & & \multicolumn{2}{c}{\sumCMOSP}   \\      \cline{2-3} \cline{5-6}
      \\[-1.8ex]
      Dec.\ Variant
      & \thetaTwoP-c & [T\ref{thm:makespan_ThetaTwoP}] &
      & NP-c &  [T\ref{thm:avg_NP}]  \\
      \midrule
      $\nbOrg$
       & DP-h/? &  [P\ref{prop:makespan_DPh_k}] & & W[1]-h/? & [P\ref{prop:avg_W1_m}]\\
      $\nbMachine$
      & W-h\jansenfollows/XP & [P\ref{prop:makespan_XP_m}] & 
      & ? & ?\\
      $\nbTask$
      & FPT & [P\ref{prop:makespan_FPT_n}] & 
      & FPT & [P\ref{prop:avg_FPT_n}]\\
      $\targetValue$
      & FPT & [C\ref{cor:makespan_FPT_target}] & 
      & FPT & [C\ref{prop:avg_FPT_target}]\\
      $\maxProcTime$
      &? & ? &
      &  ? & ?\\
      $\maxNbTask+\maxNbMachine$
       & NP-c\cohenfollows & [P\ref{prop:makespan_mmax_nmax}] &
      & NP-c & [T\ref{thm:avg_NP}]\\
      $\maxProcTime +\nbOrg$
      & FPT & [T\ref{thm:pmax+k}] & 
      &? & ?\\
      $\maxProcTime+\nbMachine$
      & FPT & [T\ref{thm:pmax+k}] & 
      & ?/XP & [P\ref{prop:avg_XP_m}]\\
      $\maxProcTime+\maxNbTask$
      & FPT & [C\ref{cor:makespan_FPT_nmax_pmax}] & 
      & ? & ?\\
    \end{tabular}}
}

\caption{See the introduction for the definition of the parameters. %
  ``DP-h'' means the problem remains DP-hard even if the value of the corresponding parameter is a constant.
  ``W-h\jansenfollows{}'' means the problem is W[1]-hard and it is due to~\protect\citeauthor{jansen2013bin}~\protect\shortcite{jansen2013bin}.
  ``NP-c'' for the parameter combination~$\maxNbTask+\maxNbMachine$ means that the decision variant is contained in NP when either of the parameters is a constant and it remains NP-hard even if both parameters have values bounded by a constant, the \makespanMOSP\ proof follows directly from~\protect\cite{cohen2011multi}. Note that all other two parameter combinations either have one parameter subsumed by the other, or have the result follow directly from another.}
\label{tab:result_table}
\end{table}

\section{Minimizing the Makespan}
\label{sec:makespan}
\appendixsection{sec:makespan}
\subsection{General Complexity}

We start this section by showing that \makespanMOSPDec\ is \thetaTwoPComplete.

\begin{restatable}[\appsymb]{theorem}{thmMakespanThetaTwoP}
	\makespanMOSPDec\ is \thetaTwoPComplete.
	\label{thm:makespan_ThetaTwoP}
\end{restatable}
\appendixproofwithstatementcontinued{thm:makespan_ThetaTwoP}{\thmMakespanThetaTwoP*}{

  We start with the containment proof.
  To this end, we introduce an intermediate scheduling NP problem for \MOSP\
  and show how to answer \makespanMOSPDec\ by making only logarithmically many calls to the NP-oracle of the newly introduced problem.

\decprob{\MOSPLS~(\MOSPLSshort)}{An instance \inst of \MOSP, two integers $\targetValue_{\glob}$ and $T'$.}{Are there two schedules~\sched and  $\sched_{\loc}$ of all jobs such that:
\begin{compactenum}[(1)]
	\item For all~$(i,j) \in [\nbOrg]\times [\nbTaskOrg{i}]\colon \machineTaskSchedule{i}{j}{\sched_{\loc}} \in \setMachineOrg{i}$,
	\item $\sum_{i \in [\nbOrg]} \left(\max_{j \in [\nbTaskOrg{i}]}\completionTimeTaskSchedule{i}{j}{\sched_{\loc}}\right) \leq T'$,
	\item $\left(\max_{\taskOrg{i}{j} \in \setTask}\completionTimeTaskSchedule{i}{j}{\sched}\right)\leq\targetValue_{\glob}$, and
	\item for all $\displaystyle(i,j)\in[\nbOrg]\times [\nbTaskOrg{i}] \colon \completionTimeTaskSchedule{i}{j}{\sched} \leq \max_{j'\in[\nbTaskOrg{i}]}\{\completionTimeTaskSchedule{i}{j'}{\sched_{\loc}}\}$?
\end{compactenum}
}

Clearly, \MOSPLSshort\ is contained in \NP\ as we can check in polynomial time
whether two given schedules~\sched and $\sched_{\loc}$  fulfill the conditions.
Intuitively, this problem asks whether there is a local schedule with sum of makespans of the organizations equal to~$T'$
and a global schedule with makespan at most~$\targetValue_{\glob}$
such that no organization has a larger makespan in the global schedule than in the local schedule.
We now describe an algorithm answering the \makespanMOSP\ problem using only a logarithmic number of calls to an oracle solving \MOSPLSshort.

Let $I$ be an instance of \makespanMOSP\ and $\targetValue$ the target makespan.
First, we perform a binary search on \MOSPLSshort\ to find the minimum sum~$T'$ of makespans among all local schedules of~$I$.
We can do this because %
if $\sched_{\loc}$ is a local schedule with minimum sum~$T'$ of makespans,
then $(I, T', T')$ is a yes-instance of \MOSPLSshort\ such that  $(\sched,\sched_{\loc})$ with $\sched=\sched_{\loc}$ is a witness.
Formally, we start with $T'=\maxProcTime\cdot\nbTask$. %
For every NP-oracle call, we set $\targetValue_{\glob}=T'$ and then do a binary search to find the minimum value~$T'$ towards which the instance is still a yes-instance of \MOSPLSshort.
Note that a local schedule with minimum sum~$T'$ of makespans among all local schedules is also
an optimal local schedule for each organization.
This is because if one organization would have a smaller local-makespan,
then by exchanging the corresponding schedule one would get a smaller sum of makespans of all organizations.

Once the minimum sum is found,
we make one last call of the \NP-oracle, where we set $T'$ to be the found minimum and $\targetValue_{\glob}=\targetValue$; recall that $\targetValue$ is the target makespan.
We answer yes if and only if the last call gives a yes-answer.
The correctness follows by checking the definition. 
This completes the containment proof.

Regarding hardness, we only give a brief sketch and defer the detailed proof to the appendix.
We reduce from a \thetaTwoPComplete\ problem consisting of comparing two ordered sets of \ThreePartition instances, where we assume that in each ordered set, all yes-instances appear before all no-instances.
An instance of the \thetaTwoPComplete\ problem is a yes-instance
if and only if there are more \ThreePartition\ yes-instances in the first set than in the second.
We group instances by pairs, one from the first set, \instTP, and one from the second, \instTPPrime, and create a set of organizations for each pair.
The local schedules of these organizations are such that if \instTPPrime is a yes-instance, then \instTP\ must also be a yes-instance to meet both individual rationality and the makespan requirement of \targetValue.  
}{

\setcounter{theorem}{1}

We now move to the hardness proof.
We start by presenting the \ThreePartition problem formally and observing a useful property.

\decprob{\ThreePartition}{An integer \sumTriplet, a set \setInt of 3\nbTriplet integers $\{\integer_1,\dots,\integer_{3\nbTriplet}\}$ such that $\sum\limits_{i \in [3\nbTriplet]} \integer_i=\nbTriplet\sumTriplet=\totsum$.}{Is there a partition of \setInt into \nbTriplet triplets $\{\triplet_1,\dots,\triplet_{\nbTriplet}\}$, such that the sum of the integers in each triplet is exactly \sumTriplet?}

The \ThreePartition problem is NP-hard~\cite{garey1979computers} even if all integers in \setInt are strictly larger than $\sumTriplet/4$ and strictly smaller than $\sumTriplet/2$ and even if the values of all integers are bounded by a polynomial of \nbTriplet. %

\begin{definition}
    Given an instance \instTP of the \ThreePartition problem and an integer $l$%
    , we denote by \instTPTimes{\instTP}{l} the instance of \ThreePartition in which each integer and the target sum of a triplet have been multiplied by $l$. In other words, if the input of \instTP is the set $\{\integer_1,\dots,\integer_{3\nbTriplet}\}$ and $\sumTriplet$ as an input, then \instTPTimes{\instTP}{l} has the set $\{\integer'_1,\dots,\integer'_{3\nbTriplet}\}$, where $\integer'_i=\integer_i \cdot l$ and $\sumTriplet'=\sumTriplet \cdot l$ as an input.%
\end{definition}

\begin{obs}
    \instTPTimes{\instTP}{l} is a yes-instance of the \ThreePartition problem if and only if \instTP is a yes-instance of the \ThreePartition problem.
    \label{obs:yesThreePartTimes}
\end{obs}

We will show \thetaTwoPHardness of \makespanMOSPDec by reduction from the \ThreePartitionComparison problem, that we define now. For an instance \instTP of \ThreePartition, we define \answer{\instTP} as $1$ if \instTP is a yes-instance and $0$ otherwise.
 
\decprob{\ThreePartitionComparison}{Two sets $\setInstOne=\{\instTP_1,\dots,\instTP_{\nbInst}\}$ and $\setInstPrime=\{\instTPPrime_1,\dots,\instTPPrime_{\nbInst'}\}$ of instances of the \ThreePartition problem.}{Are there strictly more yes-instances in \setInstOne than in \setInstPrime?}

From a direct application of Theorem 3.2 of \cite{Lukasiewicz2017}, \ThreePartitionComparison is \thetaTwoPComplete, even if $\nbInst=\nbInstPrime$ and if $\answer{\instTP_1}\geq \answer{\instTP_2} \geq \dots \geq \answer{\instTP_{\nbInst}}$ and $\answer{\instTPPrime_1}\geq \answer{\instTPPrime_2} \geq \dots \geq \answer{\instTPPrime_{\nbInst}}$, we will assume in the reduction that both these conditions are fulfilled. %

In the reduced instance, we aim at an individually rational schedule with a makespan \targetValue. To reach this makespan, it is necessary that each machine is ``perfectly" used, i.e., the total processing time of jobs assigned to it is precisely \targetValue. We will create gadgets for each pair of instances $\instTP_i$ and $\instTPPrime_{i-1}$. If $\instTPPrime_{i-1}$ is a no-instance, then regardless of $\instTP_i$, it will be possible to use these machines perfectly. If $\instTPPrime_{i-1}$ is a yes-instance, then it will only be possible to use machines perfectly if $\instTP_i$ is also a yes-instance. In the following, intuitions are written in \intuition{italic}.

For an instance \instTP of \ThreePartition, we denote by \nbTripletInst{\instTP} the number of triplets of the instance, by \sumTripletInst{\instTP} the value that each triplet is required to sum to %
, and by \integerInst{\instTP}{i} the $i^{th}$ integer in \instTP.

We start the reduction by creating a yes-instance of \ThreePartition $\instTPPrime_{0}$, with the same number of integers as $\instTP_{1}$ by choosing a target value \sumTriplet, e.g., 6, and putting only integers of value $\sumTriplet/3$. We also create a no-instance of \ThreePartition $\instTP_{\nbInst+1}$, with the same number of integers than $\instTPPrime_{\nbInst}$ by choosing a target value \sumTriplet, e.g., 18, and putting the same number of integers of value $\sumTriplet/3-1$ and $\sumTriplet/3+1$, and one of value $\sumTriplet/3$ if there is an odd number of integers in $\instTPPrime_{\nbInst}$.
\intuition{These instances are here to complete gadgets containing, respectively, \instTPInd{1} and \instTPPrimeInd{\nbInst}. Since \instTPPrimeInd{0} is a yes-instance, for the reduced instance to be a yes-instance, it is necessary that \instTPInd{1} is also a yes instance. Note that this is required for the \ThreePartitionComparison instance to be a yes-instance as well, since the question asks if there are \textit{strictly} more yes-instances in \setInstOne than in \setInstPrime. With the same idea in mind, \instTPInd{\nbInst+1} is a no-instance, therefore  for the reduced instance to be a yes-instance, it is needed that \instTPPrimeInd{\nbInst} is a no-instance, which is also required for the \ThreePartitionComparison instance to be a yes-instance.} 

For all $i \in [\nbInst+1]$, we define $\nbTripletMax{i}$ as the maximum number of triplets between $\instTP_i$ and $\instTPPrime_{i-1}$, i.e., $\nbTripletMax{i}=\max\{\nbTripletInst{\instTP_i},\nbTripletInst{\instTPPrimeInd{i-1}}\}$.

We then perform for all $i \in [\nbInst+1]$ the following modification: We replace \instTPInd{i} with \instTPTimes{\instTPInd{i}}{\sumTripletInst{\instTPPrimeInd{i-1}}} and \instTPPrimeInd{i-1} with \instTPTimes{\instTPPrimeInd{i-1}}{\sumTripletInst{\instTPInd{i}}}. Note that by \cref{obs:yesThreePartTimes}, this does not change the yes/no answer to the \ThreePartition instances and therefore it does not change the yes/no answer of the \ThreePartitionComparison instance. Note that with this modification, both instances have the same target sum for triplets, which is the product of the original target sums for triplets in both instances. From now on, we will denote by \sumTripletInd{i}, the target sum of triplets for both instances \instTPPrimeInd{i-1} and \instTPInd{i}, i.e., $\sumTripletInd{i}=\sumTripletInst{\instTP_i}\cdot\sumTripletInst{\instTPPrime_{i-1}}$. If $\sumTripletInd{i}$ is an odd integer, we multiply all integers in both instances as well as $\sumTripletInd{i}$ by two.%

We define $\lengthOffset{i}=\UBOrgTwo{i-1}+2(\sumTripletInd{i}+(\nbTripletMax{i})^2\sumTripletInd{i}^2)$ with $\UBOrgTwo{0}=2$.
\intuition{This value corresponds to an offset we will put in front of jobs from gadgets in order to make sure that gadgets do not interact with each other.}
For all $i \in [\nbInst+1]$, we define $\UBOrgTwo{i}=\lengthOffset{i}+3\sumTripletInd{i}^2\nbTripletMax{i}/2$.
\intuition{This value is an upper bound on the makespans of organizations in the $i^{th}$ gadget.}

In the reduced instance the largest integer will be of value $4\UBOrgTwo{\nbInst+1}=4\sum_{i \in [\nbInst+1]}\left(2(\sumTripletInd{i}+(\nbTripletMax{i})^2\sumTripletInd{i}^2)+3\sumTripletInd{i}^2\nbTripletMax{i}/2\right)+8$. Since the value of each integer in the \ThreePartition instances are bounded by a polynomial of the number of integers, this value is bounded by a polynomial of the total number of integers in the \ThreePartitionComparison instance.

We are now ready to move to the construction of the reduced instance.

\noindent For all $i$ in $[\nbInst+1]$, we create one \myemph{gadget} consisting of five organizations:

\begin{compactitem}
	\item $\org_{1+5(i-1)}$: This organization owns $\nbTripletMax{i}+1$ machines. It owns one ``makespan" job \taskOrg{1+5(i-1)}{0} of processing time $\lengthOffset{i}+\sumTripletInd{i}+\sumTripletInd{i}^2\nbTripletMax{i}$. It also owns a set of ``integer" jobs, denoted by \setIntegerOne{i}, containing:
	\begin{compactitem}
		\item 1 job for each integer in $\instTP_i$, i.e., for all $j \in [3\nbTripletInst{\instTP_i}]$, we create a job $\taskOrg{1+5(i-1)}{j}$ with processing time $\procTimeTask{1+5(i-1)}{j}=\integerInst{\instTP_i}{j}$ and
		\item if $\nbTripletMax{i}>\nbTripletInst{\instTP_i}$, it also owns $\nbTripletMax{i}-\nbTripletInst{\instTP_i}$ jobs $\taskOrg{1+5(i-1)}{3\nbTripletInst{\instTP_i}+1}$ to $\taskOrg{1+5(i-1)}{2\nbTripletInst{\instTP_i}+\nbTripletMax{i}}$ of processing time $\sumTripletInst{\instTP_i}$.
	\end{compactitem}		
	Note that regardless of whether $\instTP_i$ is a yes-instance or a no-instance, its optimal local makespan is necessarily $\lengthOffset{i}+\sumTripletInd{i}+\sumTripletInd{i}^2\nbTripletMax{i}$. A possible local schedule is shown in~\cref{fig:org1local}.
	\begin{figure}
	\centering
	\begin{tiny}
	\begin{tikzpicture}[scale=0.95]
		\begin{scope}[yshift=5.15cm]
			\taskColorTikz{\taskOrg{1+5(i-1)}{0}}{8}{8}{5.15}{colorOne};
		\end{scope}
		\draw [decorate, decoration = {calligraphic brace,mirror}] (-0.1,5.65) --  (-0.1,2.3);
		\node[rotate=90] at (-0.5,4.1) {$\nbTripletInst{\instTP_i}+1$ machines};
		\begin{scope}[yshift=4.6cm]
			\taskColorTikz{ }{1.2}{1.2}{6.4}{colorOne}
			\taskColorTikz{ }{0.6}{1.8}{6.4}{colorOne}
			\taskColorTikz{ }{1.2}{3}{6.4}{colorOne}
		\end{scope}
		\begin{scope}[yshift=4.05cm]
			\taskColorTikz{ }{1.45}{1.45}{5.8}{colorOne}
			\taskColorTikz{ }{0.6}{2.05}{5.8}{colorOne}
			\taskColorTikz{ }{0.95}{3}{5.8}{colorOne}
		\end{scope}

		\node at (1.5,3.75) {$\dots$};
		
		\begin{scope}[yshift=2.75cm]
			\taskColorTikz{ }{1}{1}{3}{colorOne}
			\taskColorTikz{ }{1}{2}{3}{colorOne}
			\taskColorTikz{ }{0.8}{2.8}{3}{colorOne}
		\end{scope}
		
		\begin{scope}[yshift=2.2cm]
			\taskColorTikz{ }{1.2}{1.2}{2.4}{colorOne}
			\taskColorTikz{ }{1.2}{2.4}{2.4}{colorOne}
			\taskColorTikz{ }{0.8}{3.2}{2.4}{colorOne}
		\end{scope}

		\draw [decorate, decoration = {calligraphic brace,mirror}] (-0.1,2.15) --  (-0.1,0.3);
		\node [rotate=90] at (-0.8,1.15) {$\nbTripletMax{i}\!-\!\nbTripletInst{\instTP_i}$};
		\node [rotate=90] at (-0.5,1.15) {machines};
		\begin{scope}[yshift=1.65cm]
			\taskColorTikz{\taskOrg{1+5(i-1)}{3\nbTripletInst{\instTP_i}+1}}{3}{3}{1.8}{colorOne}
		\end{scope}

		\begin{scope}[yshift=1.1cm]
			\taskColorTikz{\taskOrg{1+5(i-1)}{3\nbTripletInst{\instTP_i}+2}}{3}{3}{1.2}{colorOne}
		\end{scope}		
		
		\node at (1.5,0.9) {$\dots$};
		
		\begin{scope}[yshift=0cm]
			\taskColorTikz{\taskOrg{1+5(i-1)}{2\nbTripletInst{\instTP_i}+\nbTripletMax{i}}}{3}{3}{0}{colorOne}
		\end{scope}
		
		\draw[->] (0,0) -- (8+0.2,0);
		\draw (8,+0.1) -- (8,-0.1);
		\node at (7,-0.3){$\lengthOffset{i}+\sumTripletInd{i}+\sumTripletInd{i}^2\nbTripletMax{i}$};
		\draw (3,+0.1) -- (3,-0.1) node[below]{$\sumTripletInd{i}$};
		\draw (0,+0.1) -- (0,-0.1) node[below]{$0$};
	\end{tikzpicture}
	\end{tiny}
	\caption{Representation of the jobs and machines of organization $\org_{1+5(i-1)}$}
	\label{fig:org1local}
	\end{figure}
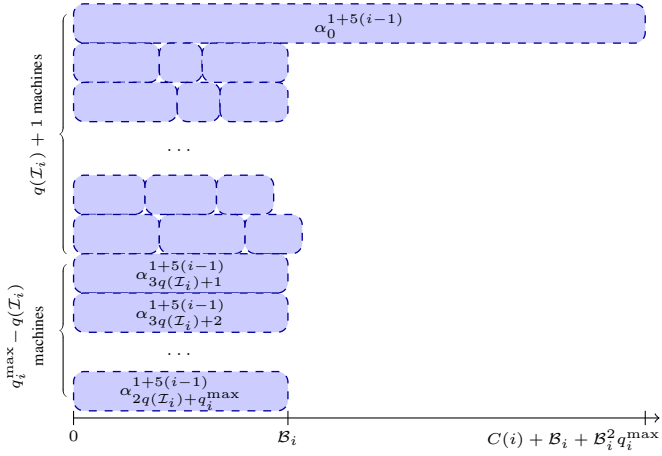

	\item $\org_{2+5(i-1)}$: This organization owns $\nbTripletMax{i}$ machines. It owns a set \setIntegerTwo{i} of ``integer" jobs which contains:
	\begin{compactitem}
		\item One job for each integer in $\instTPPrime_{i-1}$. Each of these job has processing time equal to the value of the integer multiplied by $\sumTripletInd{i}\nbTripletMax{i}$, i.e., for all $j \in [3\nbTripletInst{\instTPPrime_{i-1}}]$, we create a job $\taskOrg{2+5(i-1)}{j}$ with processing time $\procTimeTask{2+5(i-1)}{j}=\sumTripletInd{i}\nbTripletMax{i}\cdot \integerInst{\instTPPrime_{i-1}}{j}$.
		\item If $\nbTripletMax{i}>\nbTripletInst{\instTPPrime_{i-1}}$, it also owns $\nbTripletMax{i}-\nbTripletInst{\instTPPrime_{i-1}}$ jobs $\taskOrg{2+5(i-1)}{3\nbTripletInst{\instTPPrime_{i-1}}+1}$ to $\taskOrg{2+5(i-1)}{2\nbTripletInst{\instTPPrime_{i-1}}+\nbTripletMax{i}}$ of processing time $\sumTripletInd{i}\nbTripletMax{i}\times \sumTripletInd{i}$. 
	\end{compactitem}
	Organization $\org_{2+5(i-1)}$ also owns \nbTripletMax{i} ``offset "jobs of processing time \lengthOffset{i} denoted by $\taskOrg{2+5(i-1)}{2\nbTripletInst{\instTPPrime_{i-1}}+\nbTripletMax{i}+1}$ to $\taskOrg{2+5(i-1)}{2\nbTripletInst{\instTPPrime_{i-1}}+2\nbTripletMax{i}}$. A possible local schedule is shown in~\cref{fig:org2local}.

	\begin{figure}
	\centering
	\begin{tiny}
	\begin{tikzpicture}[scale=0.95]
		
		\draw [decorate, decoration = {calligraphic brace,mirror}] (-0.1,4.9) --  (-0.1,2.25);
		\node [rotate=90] at (-0.5,3.75) {$\nbTripletInst{\instTPPrime_{i-1}}$ machines};

		\begin{scope}[yshift=4.3cm]
			\taskColorTikz{ }{1}{5}{6.4}{colorTwo}
			\taskColorTikz{ }{0.6}{5.6}{6.4}{colorTwo}
			\taskColorTikz{ }{1}{6.6}{6.4}{colorTwo}
			\taskColorTikz{\taskOrg{2+5(i-1)}{2\nbTripletInst{\instTPPrime_{i-1}}+\nbTripletMax{i}+1}}{4}{4}{6.4}{colorTwo}	
		\end{scope}
		
		\begin{scope}[yshift=3.75cm]
			\taskColorTikz{ }{1.2}{5.2}{5.8}{colorTwo}
			\taskColorTikz{ }{0.6}{5.8}{5.8}{colorTwo}
			\taskColorTikz{ }{1.2}{7}{5.8}{colorTwo}
			\taskColorTikz{\taskOrg{2+5(i-1)}{2\nbTripletInst{\instTPPrime_{i-1}}+\nbTripletMax{i}+2}}{4}{4}{5.8}{colorTwo}	
		\end{scope}
		\node at (2,2.25) {$\dots$};
		\node at (5.5,3.6) {$\dots$};
		\begin{scope}[yshift=2.8cm]
			\taskColorTikz{ }{0.8}{4.8}{3}{colorTwo}
			\taskColorTikz{ }{1}{5.8}{3}{colorTwo}
			\taskColorTikz{ }{1.2}{7}{3}{colorTwo}
		\end{scope}
		\begin{scope}[yshift=2.25cm]
			\taskColorTikz{ }{1}{5}{2.4}{colorTwo}
			\taskColorTikz{ }{1.2}{6.2}{2.4}{colorTwo}
			\taskColorTikz{ }{1}{7.2}{2.4}{colorTwo}
		\end{scope}
		
		\draw [decorate, decoration = {calligraphic brace,mirror}] (-0.1,2.15) --  (-0.1,0.3);
		\node [rotate=90] at (-0.8,1.15) {$\nbTripletMax{i}\!-\!\nbTripletInst{\instTPPrime_{i-1}}$};
		\node [rotate=90] at (-0.5,1.15) {machines};

		\begin{scope}[yshift=1.65cm]
			\taskColorTikz{\taskOrg{2+5(i-1)}{3\nbTripletInst{\instTPPrime_{i-1}}+1}}{3}{7}{1.8}{colorTwo}
		\end{scope}	
		\begin{scope}[yshift=1.1cm]
			\taskColorTikz{\taskOrg{2+5(i-1)}{3\nbTripletInst{\instTPPrime_{i-1}}+2}}{3}{7}{1.2}{colorTwo}
		\end{scope}
		\node at (5.5,0.85) {$\dots$};
		\begin{scope}[yshift=0cm]
			\taskColorTikz{\taskOrg{2+5(i-1)}{2\nbTripletInst{\instTPPrime_{i-1}}+2\nbTripletMax{i}}}{4}{4}{0}{colorTwo}
			\taskColorTikz{\taskOrg{2+5(i-1)}{2\nbTripletInst{\instTPPrimeInd{i-1}}+\nbTripletMax{i}}}{3}{7}{0}{colorTwo}	
		\end{scope}

		\draw[->] (0,0) -- (8+0.2,0);
		\draw (8,+0.1) -- (8,-0.5);
		\node at (7,-0.7) {$\lengthOffset{i}+\sumTripletInd{i}^2\nbTripletMax{i}+\sumTripletInd{i}$};
		\draw (7,+0.1) -- (7,-0.1) node[below left]{$\lengthOffset{i}+\sumTripletInd{i}^2\nbTripletMax{i}$};
		\draw (4,+0.1) -- (4,-0.1) node[below]{$\lengthOffset{i}$};
		\draw (0,+0.1) -- (0,-0.1) node[below]{$0$};

	\end{tikzpicture}
	\end{tiny}
	\caption{Representation of the jobs and machines of organization $\org_{2+5(i-1)}$}
	\label{fig:org2local}
	\end{figure}
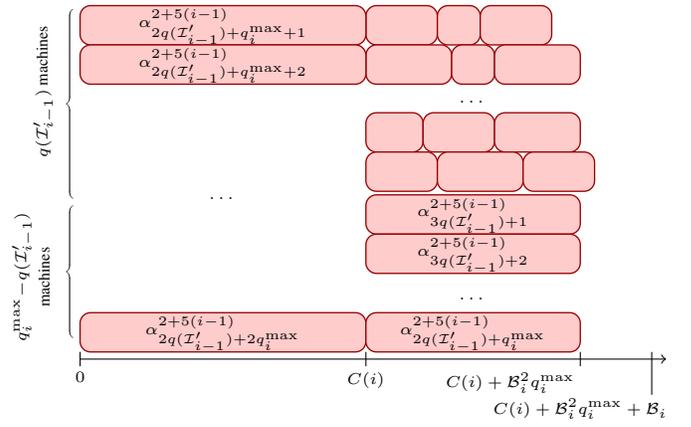

	\item $\org_{3+5(i-1)}$: This organization, also called \myemph{gadget equalizing} organization, owns 2\nbTripletMax{i} machines. It also owns \nbTripletMax{i} jobs of processing time \UBOrgTwo{i}, denoted by $\taskOrg{3+5(i-1)}{1}$ to $\taskOrg{3+5(i-1)}{\nbTripletMax{i}}$. It also owns a set denoted by \setEqualizingThree{i} of $\nbTripletMax{i}\left(\UBOrgTwo{i}-(\lengthOffset{i}+\sumTripletInd{i}+\sumTriplet{i}^2\nbTripletMax{i})\right)/\sumTripletInd{i}$ ``equalizing" jobs of processing time $\sumTripletInd{i}$. \intuition{These jobs allow us to ``fill" up to \nbTripletMax{i} machines up to $\UBOrgTwo{i}$ which will be necessary to meet the target makespan.} A possible local schedule is shown in~\cref{fig:org3local}.

\begin{figure}
	\centering
	\begin{tiny}
	\begin{tikzpicture}[scale=0.95]
		\draw [decorate, decoration = {calligraphic brace,mirror}] (-0.1,6.2) --  (-0.1,3.4);
		\node[rotate=90] at (-0.5,4.9) {$\nbTripletMax{i}$ machines};

		\begin{scope}[yshift=5.75cm]
		
			\taskColorTikz{ }{1}{1}{6.4}{colorThree}
			\node at (2.5,0.3) {$\dots$};
			\taskColorTikz{ }{1}{5}{6.4}{colorThree}
		\end{scope}
		
		\begin{scope}[yshift=5.2cm]
			\taskColorTikz{ }{1}{1}{5.8}{colorThree}
			\node at (2.5,0.3) {$\dots$};
			\taskColorTikz{ }{1}{5}{5.8}{colorThree}
		
		\end{scope}
				
		\node at (2.5,4.45) {$\dots$};
		
		\begin{scope}[yshift=3.2cm]
			\taskColorTikz{ }{1}{1}{0}{colorThree}
			\node at (2.5,0.3) {$\dots$};
			\taskColorTikz{ }{1}{5}{0}{colorThree}
		\end{scope}

		\draw [decorate, decoration = {calligraphic brace,mirror}] (-0.1,3) --  (-0.1,0);
		\node[rotate=90] at (-0.5,1.65) {$\nbTripletMax{i}$ machines};

		\begin{scope}[yshift=2.55cm]
		
			\taskColorTikz{\taskOrg{3+5(i-1)}{1}}{8}{8}{6.4}{colorThree}
		\end{scope}
		
		\begin{scope}[yshift=2cm]
			\taskColorTikz{\taskOrg{3+5(i-1)}{2}}{8}{8}{5.8}{colorThree}
		
		\end{scope}
				
		\node at (4,1.25) {$\dots$};
		
		\begin{scope}[yshift=0cm]
			\taskColorTikz{\taskOrg{3+5(i-1)}{\nbTripletMax{i}}}{8}{8}{0}{colorThree}
		\end{scope}

		\draw [->](0,0) -- (8+0.2,0);
		\draw (5,+0.1) -- (5,-0.1) node[below]{$\left(\!\sumTripletInd{i}^2\nbTripletMax{i}\!\right)\!/\!2$};
		\draw (1,+0.1) -- (1,-0.1) node[below]{$\sumTripletInd{i}$};
		\draw (8,+0.1) -- (8,-0.1) node[below left]{$\UBOrgTwo{i}$};
		\draw (0,+0.1) -- (0,-0.1) node[below]{$0$};

	\end{tikzpicture}
	\end{tiny}
	\caption{Representation of the jobs and machines of organization $\org_{3+5(i-1)}$}
	\label{fig:org3local}
	\end{figure}
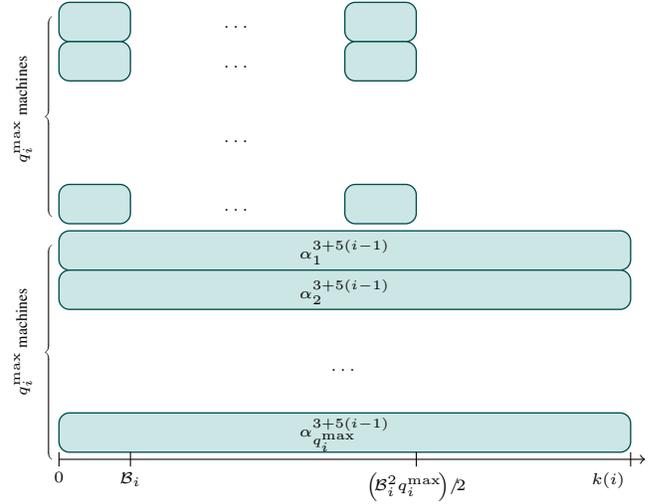

	\item $\org_{4+5(i-1)}$: This organization owns \nbTripletMax{i} machines and \nbTripletMax{i} ``offset" jobs of processing time \lengthOffset{i}, denoted by $\taskOrg{4+5(i-1)}{1}$ to $\taskOrg{4+5(i-1)}{\nbTripletMax{i}}$. A possible local schedule is shown in~\cref{fig:org4local}.
	\begin{figure}
	\centering
	\begin{tiny}
	\begin{tikzpicture}[scale=0.95]
		
		\draw [decorate, decoration = {calligraphic brace,mirror}] (-0.1,3) --  (-0.1,0);
		\node [rotate=90] at (-1,1.65) {$\nbTripletMax{i}$ machines};
		\begin{scope}[yshift=2.55cm]
			\taskColorTikz{\taskOrg{4+5(i-1)}{1}}{4}{4}{6.4}{colorFour}
		\end{scope}
		\begin{scope}[yshift=2cm]
			\taskColorTikz{\taskOrg{4+5(i-1)}{2}}{4}{4}{5.8}{colorFour}
			\end{scope}
		\node at (2,1.25) {$\dots$};
		
		\begin{scope}[yshift=0cm]
			\taskColorTikz{\taskOrg{4+5(i-1)}{\nbTripletMax{i}}}{4}{4}{0}{colorFour}	
		\end{scope}
		
		\draw[->] (0,0) -- (8+0.2,0);
		
		\draw (4,+0.1) -- (4,-0.1) node[below]{$\lengthOffset{i}$};
		\draw (0,+0.1) -- (0,-0.1) node[below]{$0$};		
	\end{tikzpicture}
	\end{tiny}
	\caption{Representation of the jobs and machines of organization $\org_{4+5(i-1)}$}
	\label{fig:org4local}
	\end{figure}
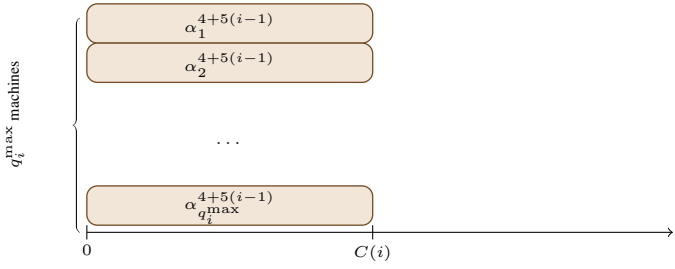
	\item $\org_{5+5(i-1)}$: This organization owns \nbTripletMax{i} machines and \nbTripletMax{i} jobs of processing time 1, denoted by $\taskOrg{5+5(i-1)}{1}$ to $\taskOrg{5+5(i-1)}{\nbTripletMax{i}}$. A possible local schedule is shown in~\cref{fig:org5local}. We note here that all jobs, except the one created for organizations $\org_{5+5(i-1)}$ are of processing time strictly larger than 1.

	\begin{figure}
	\centering
	\begin{tiny}
	\begin{tikzpicture}[scale=0.95]
		
		\draw [decorate, decoration = {calligraphic brace,mirror}] (-0.1,3) --  (-0.1,0);
		\node [rotate=90] at (-1,1.65) {$\nbTripletMax{i}$ machines};
		\begin{scope}[yshift=2.55cm]
			\taskColorTikz{\taskOrg{5+5(i-1)}{1}}{2}{2}{6.4}{colorFive}
		\end{scope}
		\begin{scope}[yshift=2cm]
			\taskColorTikz{\taskOrg{5+5(i-1)}{2}}{2}{2}{5.8}{colorFive}
		\end{scope}
		\node at (1.5,1.25) {$\dots$};
		\begin{scope}[yshift=0cm]
			\taskColorTikz{\taskOrg{5+5(i-1)}{\nbTripletMax{i}}}{2}{2}{0}{colorFive}	
		\end{scope}	
		\draw[->] (0,0) -- (8+0.2,0);
		
		\draw (2,+0.1) -- (2,-0.1) node[below]{$1$};
		\draw (0,+0.1) -- (0,-0.1) node[below]{$0$};		
	\end{tikzpicture}
	\end{tiny}
	\caption{Representation of the jobs and machines of organization $\org_{5+5(i-1)}$}
	\label{fig:org5local}
	\end{figure}
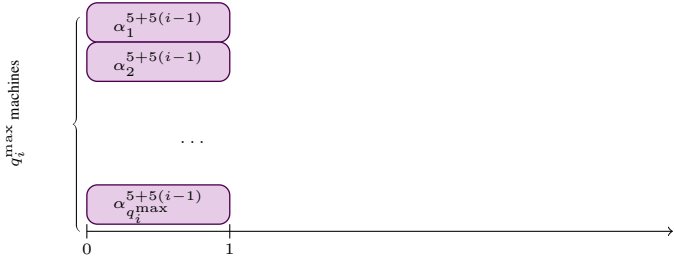
\end{compactitem}

We complete the construction by creating one \myemph{global equalizing} organization $\org_{5\nbInst+1}$ with many long jobs of specific processing time. We set $\targetValue=4(\UBOrgTwo{\nbInst+1})$. Organization $\org_{5\nbInst+1}$ owns one machine and:
\begin{compactitem}
	\item $1+\sum\limits_{i \in [\nbInst+1]} 3\nbTripletMax{i}$ jobs of processing time \targetValue,
	\item for all $i \in [\nbInst+1]$ one job of processing time $\targetValue-(\lengthOffset{i}+\sumTripletInd{i}+\sumTripletInd{i}^2\nbTripletMax{i})$,
	\item for all $i \in [\nbInst+1]$, $2\nbTripletMax{i}$ jobs of processing time $\targetValue-\UBOrgTwo{i}$, and
	\item for all $i \in [\nbInst+1]$ $\nbTripletMax{i}$ jobs of processing time $\targetValue-(\lengthOffset{i}+1)$
\end{compactitem}

Note that this organization has $1+\sum\limits_{i \in [\nbInst+1]} 3\nbTripletMax{i}+1+2\nbTripletMax{i}+\nbTripletMax{i}=1+\sum\limits_{i \in [\nbInst+1]} 1+6\nbTripletMax{i}$ jobs and that the instance has in total the same number of machines as each gadget has 5 organizations having in total $1+6\nbTripletMax{i}$ machines and the global equalizing organization has one machine.

\intuition{Each of the jobs of the global equalizing organization matches a machine from one of the gadgets. These very large jobs will leave a precise amount of time for machines to process the jobs of the gadget and this will require the machines to be used perfectly. However, if $\instTPPrime_{i-1}$ is a yes instance, then the only way to put the jobs corresponding to $\instTP_i$ on the same machines is to form triplets of the same size, which is only possible if $\instTP_i$ is also a yes instance.}

We start by a few straightforward observations regarding the reduced instance.

\begin{obs}
In an individually rational schedule with makespan \targetValue, each job of processing time \targetValue is alone on a machine.
\label{obs:targetMakespanTasks}
\end{obs}

\begin{claim}
In an individually rational schedule with makespan \targetValue, the sum of the processing times of the jobs assigned to the same machine is exactly \targetValue.
\label{claim:totalLoad}
\end{claim}

\begin{proof}\renewcommand{\qed}{\hfill (end of the proof of~\cref{claim:totalLoad})~$\diamond$}
Let us compute the total load, i.e., the sum of the processing times of all jobs. We split the computation by gadget and match the load of a gadget with corresponding jobs of the global equalizing organization. For the $i^{th}$ gadget:
\begin{compactitem}
	\item One job of processing time $(\lengthOffset{i}+\sumTripletInd{i}+\sumTripletInd{i}^2\nbTripletMax{i})$, matched by the job of processing time $\targetValue-(\lengthOffset{i}+\sumTripletInd{i}+\sumTripletInd{i}^2\nbTripletMax{i})$ owned by the global equalizing organization.
	\item jobs of total processing time $\nbTripletMax{i}\cdot(\sumTripletInd{i}+\sumTripletInd{i}^2\cdot\nbTripletMax{i})$, obtained from the integers of the two instances as well as $\nbTripletMax{i}$ jobs of processing time \lengthOffset{i} and $\nbTripletMax{i}\left(\UBOrgTwo{i}-(\lengthOffset{i}+\sumTripletInd{i}+\sumTriplet{i}^2\nbTripletMax{i})\right)/\sumTripletInd{i}$ jobs of processing time $\sumTripletInd{i}$. This whole set of jobs sum up to $\nbTripletMax{i}\cdot\UBOrgTwo{i}$ and match \nbTripletMax{i} jobs of processing time $\targetValue-\UBOrgTwo{i}$ of the global equalizing organization.
	\item \nbTripletMax{i} jobs of processing time \UBOrgTwo{i}, matching the remaining \nbTripletMax{i} jobs of processing time $\targetValue-\UBOrgTwo{i}$ of the global equalizing organization.
	\item \nbTripletMax{i} jobs of processing time \lengthOffset{i} and \nbTripletMax{i} jobs of processing time 1, matching the \nbTripletMax{i} jobs of processing time $\targetValue-(\lengthOffset{i}+1)$ of the global equalizing organization.
	\item additionally, the global equalizing organization owns $3\nbTripletMax{i}$ jobs of processing time $\targetValue$.
\end{compactitem}
That is a total of $(6\nbTripletMax{i}+1)\targetValue$ for the $i^{th}$ gadget.
We complete the proof by noting that the equalizing organization also has one machine and that there is one job of processing time \targetValue not yet accounted for. The total load is then the number of machines times \targetValue, which implies that in a schedule of makespan \targetValue, each machine needs to process jobs of total processing time \targetValue precisely.
\end{proof}

\begin{claim}
In an individually rational schedule of makespan \targetValue, each machine is assigned exactly one job from the global equalizing organization. Furthermore, this job is scheduled last on this machine.
\label{claim:equalizing_one_per_machine_and_last}
\end{claim}

\begin{proof}\renewcommand{\qed}{\hfill (end of the proof of~\cref{claim:equalizing_one_per_machine_and_last})~$\diamond$}
The first part of the statement follows directly from the fact that all jobs of the global equalizing organization has a processing time of at least $3/4$ of the target makespan. Indeed, the shortest job from the global equalizing organization has a processing time of $\targetValue-\UBOrgTwo{\nbInst+1}$, which is precisely 3/4 of \targetValue. The second part is shown by observing that no other organizations have an optimal local makespan strictly larger than $\UBOrgTwo{\nbInst+1}$, which is $1/4$ of \targetValue. This implies that if a job from another organization were to be scheduled after a job from the global equalizing organization, its completion time would be strictly larger than the optimal local makespan of this organization and the schedule would then not be individually rational.\end{proof}

\begin{claim}
If an individually rational schedule with makespan \targetValue exists, then there exists an individually rational schedule with makespan \targetValue in which, for all $i \in [\nbInst+1]$, jobs $\taskOrg{1+5(i-1)}{0}$ and the job of processing time $\targetValue-(\lengthOffset{i}+\sumTripletInd{i}+\sumTripletInd{i}^2\nbTripletMax{i})$ owned by $\org_{5\nbInst+1}$ are the only jobs on one single machine.
\label{claim:targetMakespanTasksZero}
\end{claim}

\begin{proof}\renewcommand{\qed}{\hfill (end of the proof of~\cref{claim:targetMakespanTasksZero})~$\diamond$}
Let us consider a schedule~\sched with $\makespanSched{\sched}=\targetValue$. By \cref{claim:totalLoad}, each machine processes jobs of total processing time \targetValue. Because of individual rationality, job $\taskOrg{1+5(i-1)}{0}$ has to be scheduled first. By \cref{claim:equalizing_one_per_machine_and_last}, the job of processing time $\targetValue-(\sumTripletInd{i}+\sumTripletInd{i}^2\cdot\nbTripletMax{i})$ owned by $\org_{5\nbInst+1}$ has to be scheduled last. In~\sched, since $\taskOrg{1+5(i-1)}{0}$ is scheduled first and the job of processing time $\targetValue-(\sumTripletInd{i}+\sumTripletInd{i}^2\cdot\nbTripletMax{i})$ is scheduled last on their respective machines. Assuming these machines are different, we can swap the set of jobs scheduled after $\taskOrg{1+5(i-1)}{0}$ and the job of processing time $\targetValue-(\sumTripletInd{i}+\sumTripletInd{i}^2\cdot\nbTripletMax{i})$ on $\taskOrg{1+5(i-1)}{0}$'s machine and obtain a new schedule with the same makespan and in which no job starts earlier nor later than in~\sched, this new schedule is then also individually rational.
\end{proof}
Using the same argument, we can assume that jobs of processing time $\UBOrgTwo{i}$ owned by organization $\org_{3+5(i-1)}$ are scheduled on a machine followed by a job of processing time $\targetValue-\UBOrgTwo{i}$ owned by organization $\org_{5\nbInst+1}$ for all $i \in [\nbInst]$.

\begin{claim}
If an individually rational schedule with makespan \targetValue exists, then there exists a schedule with makespan \targetValue in which, for all $i \in [\nbInst+1]$ jobs $\taskOrg{3+5(i-1)}{1}$ to $\taskOrg{3+5(i-1)}{\nbTripletMax{i}}$ are each scheduled first on a machine and a job of processing time $\targetValue-(\UBOrgTwo{i})$ owned by $\org_{5\nbInst+1}$ is scheduled afterwards on the same machine.
\label{claim:targetMakespanTasksUB}
\end{claim}

\begin{claim}
In an individually rational schedule, for all $i \in [\nbInst+1]$, all jobs of organizations $\org_{4+5(i-1)}$ and $\org_{5+5(i-1)}$ are scheduled first on a machine.
\label{claim:forced_start_small}
\end{claim}

\begin{proof}\renewcommand{\qed}{\hfill (end of the proof of~\cref{claim:forced_start_small})~$\diamond$}
This is straightforward since the optimal local makespan of organization $\org_{4+5(i-1)}$ is $\lengthOffset{i}$ and all of its jobs have processing time $\lengthOffset{i}$, so if they are delayed by at least one unit of time, individual rationality is violated.
The same argument holds for $\org_{5+5(i-1)}$ by replacing $\lengthOffset{i}$ by $1$.
\end{proof}

\begin{claim}
In an individually rational schedule of makespan \targetValue, for all $i \in [\nbInst+1]$, for all $j \in \{i+1, \dots, \nbInst+1\}$, no job owned by $\org_{1+5(i-1)}$, $\org_{2+5(i-1)}$, and $\org_{3+5(i-1)}$ can start after a job of processing time \lengthOffset{j}.
\label{claim:no_later_gadget}
\end{claim}

\begin{proof}\renewcommand{\qed}{\hfill (end of the proof of~\cref{claim:no_later_gadget})~$\diamond$}This follows directly from the fact that the local makespans of $\org_{1+5(i-1)}$, $\org_{2+5(i-1)}$ and $\org_{3+5(i-1)}$ are upper bounded by \lengthOffset{i+1} and that if $j>i$, then $\lengthOffset{j}\geq\lengthOffset{i+1}$ by definition.
\end{proof}

We are now ready to prove a claim regarding the overall structure of an existing solution if the reduced instance is a yes-instance.

\begin{claim}
If an individually rational schedule of makespan \targetValue exists, then there exists an individually rational schedule of makespan \targetValue such that, for all $i \in [\nbInst+1]$:
\begin{compactitem}
	\item All jobs of processing time \lengthOffset{i} owned by $\org_{4+5(i-1)}$ are assigned to a machine with a job of $\org_{5\nbInst+1}$ of processing time $\targetValue-\UBOrgTwo{i}$ and
	\item all jobs of processing time \lengthOffset{i} owned by $\org_{2+5(i-1)}$ are assigned to a machine with a job of $\org_{5\nbInst+1}$ of processing time $\targetValue-(\lengthOffset{i}+1)$ and a job of processing time 1.
\end{compactitem}
\label{claim:structure_sched}
\end{claim}

\begin{proof}\renewcommand{\qed}{\hfill (end of the proof of~\cref{claim:structure_sched})~$\diamond$}
We prove this by recurrence over $i$, starting with $i=\nbInst+1$ and going down. Let us assume that there exists an individually rational schedule~\sched with makespan \targetValue. By~\cref{claim:targetMakespanTasksZero}, we can assume that for all $i\in [\nbInst+1]$ jobs of processing time $\targetValue-(\lengthOffset{i}+\sumTripletInd{i}+\sumTripletInd{i}^2\nbTripletMax{i})$ owned by $\org_{5\nbInst+1}$ are assigned in~\sched to the same machine than the job of processing time $\lengthOffset{i}+\sumTripletInd{i}+\sumTripletInd{i}^2\nbTripletMax{i}$ owned by $\org_{1+5(i-1)}$. By~\cref{claim:targetMakespanTasksUB}, we can assume for all $i \in [\nbInst+1]$ that $\nbTripletMax{i}$ jobs of processing time $\targetValue-\UBOrgTwo{i}$ owned by $\org_{5\nbInst+1}$ are assigned in~\sched to $\nbTripletMax{i}$ different machines, each with one of the $\nbTripletMax{i}$ jobs of processing time $\UBOrgTwo{i}$ owned by $\org_{3+5(i-1)}$. 

\noindent\textbf{Base case: $i=\nbInst+1$.} By~\cref{claim:equalizing_one_per_machine_and_last}, we know that in~\sched there is exactly one job of organization $\org_{5\nbInst+1}$ on each machine. Now observe that $\lengthOffset{\nbInst+1}>\UBOrgTwo{\nbInst}>\lengthOffset{\nbInst}+1>\dots >\UBOrgTwo{1}>\lengthOffset{1}+1$. This implies that if a job of processing time \lengthOffset{\nbInst+1} is assigned to a machine, the only jobs of $\org_{5\nbInst+1}$ that can be assigned to the same machine in~\sched are jobs of processing time either $\targetValue-(\lengthOffset{\nbInst+1}+\sumTripletInd{\nbInst+1}+\sumTripletInd{\nbInst+1}^2\nbTripletMax{\nbInst+1})$, $\targetValue-\UBOrgTwo{\nbInst+1}$, or $\targetValue-(\lengthOffset{\nbInst+1}+1)$. %
Due to the earlier assumption the job of processing time $\targetValue-(\lengthOffset{\nbInst+1}+\sumTripletInd{\nbInst+1}+\sumTripletInd{\nbInst+1}^2\nbTripletMax{\nbInst+1})$ is assigned to another machine.
Furthermore, since $2\lengthOffset{\nbInst+1}>\UBOrgTwo{\nbInst+1}>\lengthOffset{\nbInst+1}+1$ it is impossible for two jobs of processing time \lengthOffset{\nbInst+1} to be assigned to the same machine with a job of $\org_{5\nbInst+1}$. Therefore, each job of processing time \lengthOffset{\nbInst+1} has to be assigned to a distinct machine and the job of $\org_{5\nbInst+1}$ assigned to these machines are of processing time either $\targetValue-\UBOrgTwo{\nbInst+1}$ or $\targetValue-(\lengthOffset{\nbInst+1}+1)$. Now observe that all jobs of processing time exactly $1$ is owned by an organization with optimal local makespan of 1. This means that jobs of processing time 1 cannot be schedule after any job of processing time \lengthOffset{\nbInst+1}, furthermore, jobs of processing time \lengthOffset{\nbInst+1} owned by $\org_{5{\nbInst-1}+4}$ have to be scheduled first in~\sched by \cref{claim:forced_start_small}. This implies that jobs with processing time $\targetValue-(\lengthOffset{\nbInst+1}+1)$ are necessarily assigned to machines to which a job of processing time \lengthOffset{\nbInst+1} owned by $\org_{2+5{\nbInst}}$ is assigned. Furthermore this machine necessarily starts with a job of processing time 1, then the job of processing time \lengthOffset{\nbInst+1} and then the job owned by the global equalizing organization.

\noindent\textbf{Recursion.} Assume that for all $j \in \{i+1, \dots, \nbInst+1\}$, jobs of processing time $\targetValue-\UBOrgTwo{j}$ and $\targetValue-(\lengthOffset{j}+1)$ are assigned to machine with jobs of processing time $\lengthOffset{j}$ owned by organizations $\org_{5(j-1)+2}$ and $\org_{5(j-1)+4}$. By~\cref{claim:no_later_gadget,claim:forced_start_small}, for all $j \in \{i+1, \dots, \nbInst\}$ jobs of processing time \lengthOffset{i} cannot start after jobs of processing time \lengthOffset{j}. Therefore, for all $j \in \{i+1, \dots, \nbInst\}$ no job of processing time \lengthOffset{i} can be assigned to the same machine than jobs of processing time $\targetValue-\UBOrgTwo{j}$ and $\targetValue-(\lengthOffset{j}+1)$. Now observe that $\lengthOffset{i}>\UBOrgTwo{i-1}>\lengthOffset{i-2}+1>\dots >\UBOrgTwo{1}>\lengthOffset{1}+1$. This implies that if a job of processing time \lengthOffset{i} is assigned to a machine, the only jobs of $\org_{5\nbInst+1}$ that can be assigned to the same machine in~\sched are jobs of processing time either $\targetValue-(\lengthOffset{i}+\sumTripletInd{i}+\sumTripletInd{i}^2\nbTripletMax{i})$, $\targetValue-\UBOrgTwo{i}$, or $\targetValue-(\lengthOffset{i}+1)$. Due to the earlier assumption the job of processing time $\targetValue-(\lengthOffset{i}+\sumTripletInd{i}+\sumTripletInd{i}^2\nbTripletMax{i})$ is assigned to another machine. The rest of the proof is similar to the base case.
\end{proof}

We finally note that, in an individually rational schedule of makespan \targetValue, because of \cref{claim:no_later_gadget} and \cref{claim:structure_sched}, it follows that the integer jobs of $\org_{1+5(i-1)}$ and $\org_{2+5(i-1))}$ as well as the equalizing jobs of $\org_{3+5(i-1)}$ have to be assigned to the same machines than the jobs of $\org_{4+5(i-1)}$.

\begin{claim}
If an individually rational schedule of makespan \targetValue, then  there exists an individually rational schedule of makespan \targetValue such that for all $i \in [\nbInst+1]$ the jobs in \setIntegerOne{i} owned by $\org_{1+5(i-1)}$, in \setIntegerTwo{i} owned by $\org_{2+5(i-1)}$, and \setEqualizingThree{i} owned by $\org_{3+5(i-1)}$ are scheduled after jobs of processing time \lengthOffset{i} owned by $\org_{4+5(i-1)}$ on the same machines and before time $\UBOrgTwo{i}$.
\label{claim:final_structure}
\end{claim}

\begin{proof}\renewcommand{\qed}{\hfill (end of the proof of~\cref{claim:final_structure})~$\diamond$}
Let us assume that an individually rational schedule of makespan \targetValue exists.
We consider an individually rational schedule~\sched of makespan \targetValue such that:
\begin{compactitem}
	\item All jobs of processing time $\targetValue$ are processed alone on a machine,
	\item for all $i \in [\nbInst+1]$ jobs $\taskOrg{1+5(i-1)}{0}$ and the job of processing time $\targetValue-(\lengthOffset{i}+\sumTripletInd{i}+\sumTripletInd{i}^2\nbTripletMax{i})$ owned by $\org_{5\nbInst+1}$ are the only jobs on one single machine,
	\item for all $i \in [\nbInst+1]$ jobs $\taskOrg{3+5(i-1)}{1}$ to $\taskOrg{3+5(i-1)}{\nbTripletMax{i}}$ are each scheduled first on a machine and a job of processing time $\targetValue-(\UBOrgTwo{i})$ owned by $\org_{5\nbInst+1}$ is scheduled afterwards on the same machine,
	\item All jobs of processing time \lengthOffset{i} owned by $\org_{4+5(i-1)}$ are assigned to a machine with a job of $\org_{5\nbInst+1}$ of processing time $\targetValue-\UBOrgTwo{i}$, and
	\item all jobs of processing time \lengthOffset{i} owned by $\org_{2+5(i-1)}$ are assigned to a machine with a job of $\org_{5\nbInst+1}$ of processing time $\targetValue-(\lengthOffset{i}+1)$ and a job of processing time 1.
\end{compactitem}
By \cref{claim:targetMakespanTasksUB,claim:targetMakespanTasksZero} and \cref{claim:structure_sched}, we know that if the reduced instance is a yes instance then such a schedule exists.

By~\cref{claim:no_later_gadget}, jobs owned by $\org_{1}$, $\org_{2}$, and $\org_{3}$ and that are not assigned according to the previous assumptions, i.e., jobs from \setIntegerOne{i},\setIntegerTwo{i}, and \setEqualizingThree{i}, cannot be scheduled on any machine except the one with starting jobs of processing time \lengthOffset{1} owned by $\org_{4+5(i-1)}$ as any other machine either processes a job of processing time \lengthOffset{i}, with $i>1$ first or is already assigned a total processing time of \targetValue. Additionally, the total processing time of these jobs is of $\nbTripletMax{1}(\sumTripletInd{1}+\sumTripletInd{1}^2\nbTripletMax{1})+\nbTripletMax{1}(\UBOrgTwo{1}-(\lengthOffset{1}+\sumTripletInd{1}+\sumTripletInd{1}^2\nbTripletMax{1})=\nbTripletMax{1}(\UBOrgTwo{1}-\lengthOffset{1})$. This corresponds to the amount of time between the end of the jobs of processing time \lengthOffset{i} and the beginning of the job owned by the global equalizing organization which is of processing time $\targetValue-\UBOrgTwo{1}$. This means that no other job can be processed on these machines. 

By repeating this observation iteratively for all $i$ from 1 to $\nbInst+1$ we obtain the claim.
\end{proof}

\cref{fig:structure_opt} shows the overall structure described by \cref{obs:targetMakespanTasks,claim:targetMakespanTasksUB,claim:targetMakespanTasksZero,claim:structure_sched,claim:final_structure}.

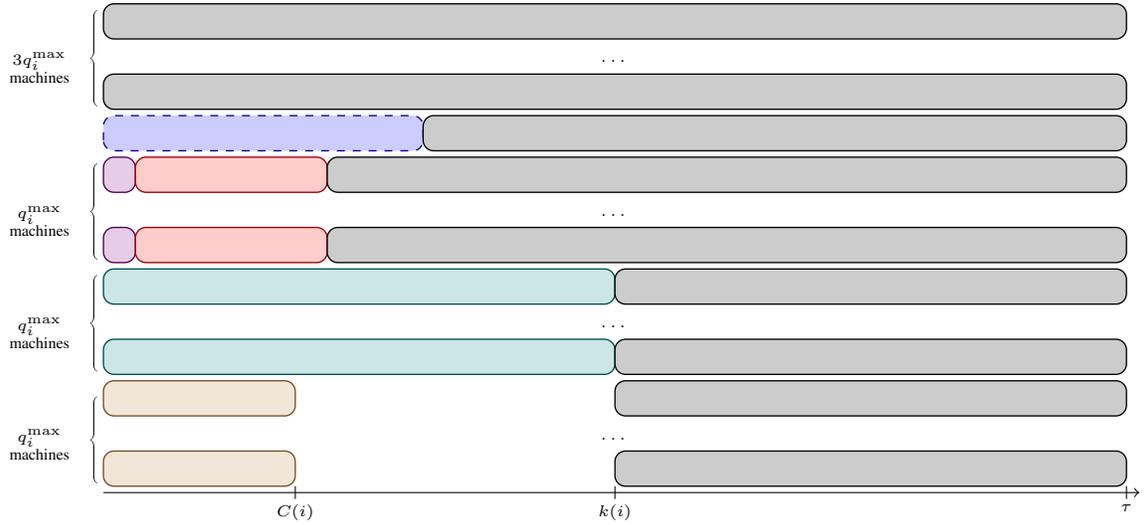
\begin{figure*}

\centering
	\begin{tiny}
	\begin{tikzpicture}[scale=0.85]
	
		\draw [decorate, decoration = {calligraphic brace,mirror}] (-0.1,7.55) --  (-0.1,6.05);
		\node at (-1,6.75) {$3\nbTripletMax{i}$};
		\node at (-1,6.5) {machines};
		\begin{scope}[yshift=7cm]
			\taskColorTikz{}{16}{16}{0}{colorEqualizing}
		\end{scope}
		\node at (8,6.75) {$\dots$};
		\begin{scope}[yshift=5.9cm]
			\taskColorTikz{}{16}{16}{0}{colorEqualizing}
		\end{scope}

		\begin{scope}[yshift=5.25cm]
			\taskColorTikz{}{5}{5}{0}{colorOne}
			\taskColorTikz{}{11}{16}{0}{colorEqualizing}
		\end{scope}
		
		\draw [decorate, decoration = {calligraphic brace,mirror}] (-0.1,5.15) --  (-0.1,3.65);
		\node at (-1,4.35) {$\nbTripletMax{i}$};
		\node at (-1,4.1) {machines};
		\begin{scope}[yshift=4.6cm]
			\taskColorTikz{}{0.5}{0.5}{0}{colorFive}
			\taskColorTikz{}{3}{3.5}{0}{colorTwo}
			\taskColorTikz{}{12.5}{16}{0}{colorEqualizing}
		\end{scope}
		\node at (8,4.35) {$\dots$};
		\begin{scope}[yshift=3.5cm]
			\taskColorTikz{}{0.5}{0.5}{0}{colorFive}
			\taskColorTikz{}{3}{3.5}{0}{colorTwo}
			\taskColorTikz{}{12.5}{16}{0}{colorEqualizing}
		\end{scope}	
	
		\draw [decorate, decoration = {calligraphic brace,mirror}] (-0.1,3.4) --  (-0.1,1.9);
		\node at (-1,2.6) {$\nbTripletMax{i}$};
		\node at (-1,2.35) {machines};
		\begin{scope}[yshift=2.85cm]
			\taskColorTikz{}{8}{8}{0}{colorThree}
			\taskColorTikz{}{8}{16}{0}{colorEqualizing}
		\end{scope}
		\node at (8,2.6) {$\dots$};
		\begin{scope}[yshift=1.75cm]
			\taskColorTikz{}{8}{8}{0}{colorThree}
			\taskColorTikz{}{8}{16}{0}{colorEqualizing}	
		\end{scope}

		\draw [decorate, decoration = {calligraphic brace,mirror}] (-0.1,1.5) --  (-0.1,0.15);
		\node at (-1,0.85) {$\nbTripletMax{i}$};
		\node at (-1,0.6) {machines};
		\begin{scope}[yshift=1.1cm]
			\taskColorTikz{}{3}{3}{0}{colorFour}
			\taskColorTikz{}{8}{16}{0}{colorEqualizing}
		\end{scope}
		\node at (8,0.85) {$\dots$};
		\begin{scope}[yshift=0cm]
			\taskColorTikz{}{3}{3}{0}{colorFour}
			\taskColorTikz{}{8}{16}{0}{colorEqualizing}	
		\end{scope}

		\draw[->] (0,0) -- (16+0.2,0);
		
		\draw (16,+0.1) -- (16,-0.1) node[below]{$\targetValue$};
		\draw (8,+0.1) -- (8,-0.1) node[below]{$\UBOrgTwo{i}$};
		\draw (3,+0.1) -- (3,-0.1) node[below]{$\lengthOffset{i}$};		
	\end{tikzpicture}
	\end{tiny}
\caption{Structure of an existing individually rational schedule with makespan \targetValue, relative to the $i^{th}$ gadget, if one exists. The space left blank in the bottom machines correspond to the time in which jobs from the sets \setIntegerOne{i},\setIntegerTwo{i} and \setEqualizingThree{i} have to be scheduled. jobs in black are owned by the global equalizing organization.}
\label{fig:structure_opt}
\end{figure*}

We will now show that the reduced instance is a yes-instance if and only if the \ThreePartitionComparison instance is a yes-instance.

\onlyifDirection To show that the reduced instance is a yes-instance only if the \ThreePartitionComparison instance is a yes-instance we prove a claim.

\begin{claim}
If the reduced instance is a yes-instance, then for all $i \in [\nbInst+1]$, if $\instTPPrime_{i-1}$ is a yes instance, then $\instTP_{i}$ is also a yes-instance.
\label{claim:validScheduleAndYesImpliesYes}
\end{claim}

\begin{proof}\renewcommand{\qed}{\hfill (end of the proof of~\cref{claim:validScheduleAndYesImpliesYes})~$\diamond$}
Let us assume that the reduced instance is a yes-instance, i.e., there exists an individually rational schedule of makespan \targetValue. Let us consider for the sake of contradiction that there exists an $i \in [\nbInst+1]$ such that $\instTPPrime_{i-1}$ is a yes instance and $\instTP_{i}$ is a no-instance. 

Let us call~\sched an individually rational schedule in which:
\begin{compactitem}
	\item All jobs of processing time $\targetValue$ are processed alone on a machine,
	\item for all $i \in [\nbInst+1]$ jobs $\taskOrg{1+5(i-1)}{0}$ and the job of processing time $\targetValue-(\lengthOffset{i}+\sumTripletInd{i}+\sumTripletInd{i}^2\nbTripletMax{i})$ owned by $\org_{5\nbInst+1}$ are the only jobs on one single machine,
	\item for all $i \in [\nbInst+1]$ jobs $\taskOrg{3+5(i-1)}{1}$ to $\taskOrg{3+5(i-1)}{\nbTripletMax{i}}$ are each scheduled first on a machine and a job of processing time $\targetValue-(\UBOrgTwo{i})$ owned by $\org_{5\nbInst+1}$ is scheduled afterwards on the same machine,
	\item all jobs of processing time \lengthOffset{i} owned by $\org_{4+5(i-1)}$ are assigned to a machine with a job of $\org_{5\nbInst+1}$ of processing time $\targetValue-\UBOrgTwo{i}$, 
	\item all jobs of processing time \lengthOffset{i} owned by $\org_{2+5(i-1)}$ are assigned to a machine with a job of $\org_{5\nbInst+1}$ of processing time $\targetValue-(\lengthOffset{i}+1)$ and a job of processing time 1, and
	\item for all $i \in [\nbInst+1]$ the jobs owned by $\org_{1+5(i-1)}$, $\org_{2+5(i-1)}$, and $\org_{3+5(i-1)}$ except the job $\taskOrg{1+5(i-1)}{0}$, the jobs of processing time \lengthOffset{i} owned by $\org_{2+5(i-1)}$ and the jobs of processing time \UBOrgTwo{i} owned by $\org_{3+5(i-1)}$ are scheduled after jobs of processing time \lengthOffset{i} owned by $\org_{4+5(i-1)}$ on the same machines and before time $\UBOrgTwo{i}$.
\end{compactitem}
By~\cref{claim:targetMakespanTasksUB,claim:targetMakespanTasksZero,claim:structure_sched,claim:final_structure} and~\cref{obs:targetMakespanTasks}, we know that if the reduced instance is a yes instance then such a schedule exists.

Now, since $\instTPPrime_{i-1}$ is a yes-instance, it means the optimal local makespan of $\org_{2+5(i-1)}$ is precisely $\lengthOffset{i}+\sumTripletInd{i}^2\nbTripletMax{i}$. Indeed, a schedule with such a makespan can be obtained by putting one job of processing time \lengthOffset{i} o each machine and dividing the jobs from \setIntegerTwo{i} according to a \ThreePartition of the corresponding integers in $\instTPPrime_{i-1}$. Since~\sched is individually rational, all jobs scheduled owned by $\org_{2+5(i-1)}$ and scheduled on the machines starting with the jobs of processing time \lengthOffset{i} of $\org_{4+5(i-1)}$ are completed by time $\lengthOffset{i}+\sumTripletInd{i}^2\nbTripletMax{i}$. As the total completion time of these jobs is of $\nbTripletMax{i} \cdot \sumTripletInd{i}^2\nbTripletMax{i}$, this means that no other job can be processed on these machines before time $\lengthOffset{i}+\sumTripletInd{i}^2\nbTripletMax{i}$. Now, because of individual rationality, each job of $\org_{1+5(i-1)}$ is completed in~\sched at the latest at time $\lengthOffset{i}+\sumTripletInd{i}+\sumTripletInd{i}^2\nbTripletMax{i}$, this means that all jobs of \setIntegerOne{i} are processed between time $\lengthOffset{i}+\sumTripletInd{i}^2\nbTripletMax{i}$ and $\lengthOffset{i}+\sumTripletInd{i}^2\nbTripletMax{i}+\sumTripletInd{i}$. This means that each machine processes these jobs for precisely $\sumTripletInd{i}$ units of time. Since jobs owned by $\org_{1+5(i-1)}$ are either of processing time \sumTripletInd{i} (if $\instTPPrime_{i-1}$ contains more integers than $\instTP_{i}$), or of processing time matching the value of the integers of $\instTP_{i}$, such a schedule can only be obtained if each machine processes either one job of processing time \sumTripletInd{i} or three jobs with total processing time \sumTripletInd{i}. It is easy to see that the integers corresponding to the sets of jobs processed by the same machine sum to \sumTripletInd{i} and therefore, the corresponding integers form valid triplets for the instance. Since all jobs can be grouped into triples of total processing time \sumTripletInd{i}, this means that all integers of $\instTP_{i}$ can be partitioned into valid triplets and $\instTP_{i}$ is then a yes-instance, a contradiction.
\end{proof}

Since $\instTPPrime_0$ is a yes-instance, for the reduced instance to be a yes-instance it is required that $\instTP_1$ is a yes instance, any additional yes-instance in $\setInstPrime$ requires an additional yes-instance in $\setInstOne$ for the reduced instance to be a yes-instance. Finally, since $\instTP_{\nbInst+1}$ is a no-instance, if all instances in $\setInstPrime$ are yes-instance, then it is impossible for the reduced instance to be a yes-instance. This completes the proof of the \onlyifDirection direction.

\ifDirection We now assume that the \ThreePartitionComparison instance is a yes-instance. This implies that there is a $j \in [\nbInst]$ such that $\forall j' \in \{0, \dots, j-1\}, \instTPPrime_{j'}$ is a yes-instance, $\forall i \in [j], \instTP_{i}$ is a yes-instance, and $\instTPPrime_{j}$ is a no-instance.

By~\cref{claim:targetMakespanTasksZero,claim:targetMakespanTasksUB,claim:structure_sched,claim:final_structure}, we know that if an individually rational schedule with makespan \targetValue exists, i.e., the reduced instance is a yes-instance, then one exists in which:
\begin{compactitem}
	\item All jobs of processing time $\targetValue$ are processed alone on a machine,
	\item for all $i \in [\nbInst+1]$ jobs $\taskOrg{1+5(i-1)}{0}$ and the job of processing time $\targetValue-(\lengthOffset{i}+\sumTripletInd{i}+\sumTripletInd{i}^2\nbTripletMax{i})$ owned by $\org_{5\nbInst+1}$ are the only jobs on one single machine,
	\item for all $i \in [\nbInst+1]$ jobs $\taskOrg{3+5(i-1)}{1}$ to $\taskOrg{3+5(i-1)}{\nbTripletMax{i}}$ are each scheduled first on a machine and a job of processing time $\targetValue-(\UBOrgTwo{i})$ owned by $\org_{5\nbInst+1}$ is scheduled afterwards on the same machine,
	\item all jobs of processing time \lengthOffset{i} owned by $\org_{4+5(i-1)}$ are assigned to a machine with a job of $\org_{5\nbInst+1}$ of processing time $\targetValue-\UBOrgTwo{i}$, 
	\item all jobs of processing time \lengthOffset{i} owned by $\org_{2+5(i-1)}$ are assigned to a machine with a job of $\org_{5\nbInst+1}$ of processing time $\targetValue-(\lengthOffset{i}+1)$ and a job of processing time 1, and
	\item for all $i \in [\nbInst+1]$ the jobs owned by $\org_{1+5(i-1)}$, $\org_{2+5(i-1)}$, and $\org_{3+5(i-1)}$ except the job $\taskOrg{1+5(i-1)}{0}$, the jobs of processing time \lengthOffset{i} owned by $\org_{2+5(i-1)}$ and the jobs of processing time \UBOrgTwo{i} owned by $\org_{3+5(i-1)}$ are scheduled after jobs of processing time \lengthOffset{i} owned by $\org_{4+5(i-1)}$ on the same machines and before time $\UBOrgTwo{i}$.
\end{compactitem}
We therefore start building a schedule~\sched following the first five properties above and provide details regarding the sixth. If no schedule fulfilling these properties exists, then the reduced instance is a no-instance. 
One can see that each machine has a total load of precisely \targetValue, assuming that the jobs from \setIntegerOne{i},\setIntegerTwo{i}, and \setEqualizingThree{i} can all be scheduled between time \lengthOffset{i} and time \UBOrgTwo{i} on machines processing jobs from $\org_{4+5(i-1)}$.\\

For all $i \in [j]$,  both \instTPPrime{i-1} and \instTP{i} are yes-instances, therefore the optimal local makespan of $\org_{2+5(i-1)}$ is $\lengthOffset{i}+\sumTripletInd{i}^2\nbTripletMax{i}$.
\begin{compactitem}
	\item jobs from \setIntegerTwo{i} owned by $\org_{2+5(i-1)}$ are scheduled in~\sched in the same way than in the local schedule, i.e., the jobs processed by the same machine keep their starting and completion time and are moved to the same machine processing the job of processing time \lengthOffset{i} owned by $\org_{4+5(i-1)}$. Note that this does not increase the makespan of $\org_{2+5(i-1)}$ in comparison to its local makespan, therefore it does not violate individual rationality.
	\item Since $\instTP_{i}$ is a yes-instance, there exists a partition of its integers into triplets such that each triplet sums to precisely \sumTripletInd{i}. By construction $\org_{1+5(i-1)}$ owns a job of processing time $\integerInst{\instTP{i}}{l}$ for all $l \in 3\nbTripletInst{\instTP{i}}$, we schedule the jobs with the corresponding processing time on one of the machines starting with a job of processing time \lengthOffset{i} owned by $\org_{4+5(i-1)}$. Each of these set of jobs starts at time $\lengthOffset{i}+\sumTripletInd{i}^2\nbTripletMax{i}$ and completes at time $\lengthOffset{i}+\sumTripletInd{i}^2\nbTripletMax{i}+\sumTripletInd{i}$. If $\nbTripletInst{\instTP{i}}>\nbTripletInst{\instTPPrime{i-1}}$, we schedule the remaining $\nbTripletMax{i}-\nbTripletInst{\instTP{i}}$ jobs of processing time \sumTripletInd{i}, if any, on the remaining machines during the same time window. Note that all these jobs are completed by time $\lengthOffset{i}+\sumTripletInd{i}^2\nbTripletMax{i}+\sumTripletInd{i}$, which is the optimal local makespan for $\org_{1+5(i-1)}$, and therefore, this does not violate individual rationality.
	\item Finally, we schedule the $\nbTripletMax{i}\left(\UBOrgTwo{i}-(\lengthOffset{i}+\sumTripletInd{i}+\sumTriplet{i}^2\nbTripletMax{i})\right)/\sumTripletInd{i}$ jobs of processing time $\sumTripletInd{i}$ of \setEqualizingThree{i} owned by $\org_{3+5(i-1)}$ on the \nbTripletMax{i} machines starting with a job of $\org_{4+5(i-1)}$ from time $\lengthOffset{i}+\sumTripletInd{i}^2\nbTripletMax{i}+\sumTripletInd{i}$ to time \UBOrgTwo{i}. Note that $\UBOrgTwo{i}-\lengthOffset{i}=3/2\sumTripletInd{i}^2\nbTripletMax{i}$, which is a multiple of \sumTripletInd{i}. This means that it is possible to fill completely the time on all machines with jobs of processing time \sumTripletInd{i}. All these jobs complete at the latest at \UBOrgTwo{i}, which is the optimal local makespan for $\org_{3+5(i-1)}$.
\end{compactitem}

For all $i \in \{j+1,\dots, \nbInst+1\}$, $\instTPPrime_{i-1}$ is a no-instance. 

We start by showing a few properties of an optimal local schedule.

\begin{claim}
For all $i\in \{j+1,\dots, \nbInst+1\}$, in an optimal local schedule for $\org_{2+5(i-1)}$, there is exactly one job of processing time \lengthOffset{i} on each machine.
\label{obs:local_one_long_task_per_machine}
\end{claim}

\begin{proof}\renewcommand{\qed}{\hfill (end of the proof of~\cref{obs:local_one_long_task_per_machine})~$\diamond$}
Let us assume that a machine processes two jobs of processing time \lengthOffset{i} in an optimal local schedule, by pigeonhole principle, there is one machine which does not process any. Even if one machine not processing any job of processing time \lengthOffset{i} processes all other jobs, its total load is $\sumTripletInd{i}^2(\nbTripletMax{i})^2$, which is strictly smaller than \lengthOffset{i}. This means that swapping one of the jobs of processing time \lengthOffset{i} from the machine with two with the load of the machine with none would reduce the makespan and the schedule would then not be optimal.
\end{proof}

Without loss of generality, we will suppose that the jobs of processing time \lengthOffset{i} are scheduled first on each machine in an optimal local schedule.

\begin{claim}
For all $i\in \{j+1,\dots, \nbInst+1\}$, in an optimal local schedule for $\org_{2+5(i-1)}$, one machine has a total load of at most $\lengthOffset{i}+\sumTripletInd{i}^2\nbTripletMax{i}-\sumTripletInd{i}\nbTripletMax{i}$.
\label{claim:local_gap}
\end{claim}

\begin{proof}\renewcommand{\qed}{\hfill (end of the proof of~\cref{claim:local_gap})~$\diamond$}
Since $\instTPPrime_{i}$ is a no-instance, it is not possible to partition its integers into triplets of sum $\sumTripletInd{i}^2\nbTripletMax{i}$. Now observe that the processing times of all the jobs of $\org_{2+5(i-1)}$, except the jobs of processing time \lengthOffset{i}, are obtained by multiplying integers by $\sumTripletInd{i}\nbTripletMax{i}$. This means that the load of each machine in an optimal local schedule for $\org_{2+5(i-1)}$ is of $\lengthOffset{i}+\sumTripletInd{i}\nbTripletMax{i}\cdot x$ where $x$ is an integer. Since $\instTPPrime_{i}$ is a no-instance it is impossible to partition the jobs of $\org_{2+5(i-1)}$ built from the integers of $\instTPPrime_{i}$ into \nbTripletMax{i} triplets of total processing time $\sumTripletInd{i}^2\nbTripletMax{i}$ as otherwise it would be possible to find triplets of integers in \instTPPrime{i} summing to \sumTripletInd{i}. This means that at least one machine is assigned a set of jobs which total processing time is strictly lower than $\lengthOffset{i}+\sumTripletInd{i}^2\nbTripletMax{i}$, and since its load is of $\lengthOffset{i}+\sumTripletInd{i}\nbTripletMax{i}\cdot x$ where $x$ is an integer, its load is of at most $\lengthOffset{i}+\sumTripletInd{i}^2\nbTripletMax{i}-\sumTripletInd{i}\nbTripletMax{i}$.
\end{proof}

\begin{claim}
For all $i\in \{j+1,\dots, \nbInst+1\}$, in an optimal local schedule for $\org_{2+5(i-1)}$, the makespan is at most \UBOrgTwo{i}.
\label{claim:UB}
\end{claim}

\begin{proof}\renewcommand{\qed}{\hfill (end of the proof of~\cref{claim:UB})~$\diamond$}
Let us consider an optimal local schedule for $\org_{2+5(i-1)}$ such that there is exactly one job of processing time \lengthOffset{i} assigned on each machine and processed first and such that its total load is strictly greater than $\UBOrgTwo{i}=\lengthOffset{i}+3/2\sumTripletInd{i}^2\nbTripletMax{i}$. Let us consider the last job to be processed by this machine. Since the value of each integer in $\instTPPrime_{i-1}$ is strictly less than $\sumTripletInst{\instTPPrime{i-1}}/2$, this job is of processing time strictly less than $\sumTripletInd{i}^2\nbTripletMax{i}/2$, which means that it starts at time at least $\lengthOffset{i}+\sumTripletInd{i}^2\nbTripletMax{i}+1$. Now, by~\cref{claim:local_gap}, one machine in the considered local schedule has a total load strictly lower than $\lengthOffset{i}+\sumTripletInd{i}^2\nbTripletMax{i}$ which means that moving the last job to this machine would lower the makespan of the schedule, contradicting its optimality.
\end{proof}

We are now ready to build the second part of the schedule~$\sched$.

For all $i\in \{j+1,\dots, \nbInst+1\}$:
\begin{compactitem}
	\item jobs from \setIntegerTwo{i} owned by $\org_{2+5(i-1)}$ are scheduled in~\sched in the same way than in the local schedule, i.e., the jobs processed by the same machine keep their starting and completion time and are moved to the machine processing the job of processing time \lengthOffset{i} owned by $\org_{4+5(i-1)}$. Note that this does not increase the makespan of $\org_{2+5(i-1)}$ in comparison to its local makespan, therefore individual rationality is not violated.
	\item By~\cref{claim:local_gap}, there exists one machine starting with a job from $\org_{4+5(i-1)}$ which processes job from $\org_{2+5(i-1)}$ to time at most $\lengthOffset{i}+\sumTripletInd{i}^2\nbTripletMax{i}-\sumTripletInd{i}\nbTripletMax{i}$, we schedule in~\sched all jobs in \setIntegerOne{i} from $\org_{1+5(i-1)}$ on this machine. As the total processing time of these jobs is $\sumTripletInd{i}\nbTripletMax{i}$, they are all completed before $\lengthOffset{i}+\sumTripletInd{i}^2\nbTripletMax{i}$, which is lower than the optimal local makespan of the organization, ensuring individual rationality for the organization.
	\item Finally, we schedule the $\nbTripletMax{i}\left(\UBOrgTwo{i}-(\lengthOffset{i}+\sumTripletInd{i}+\sumTriplet{i}^2\nbTripletMax{i})\right)/\sumTripletInd{i}$ jobs of processing time $\sumTripletInd{i}$ of $\org_{3+5(i-1)}$ on the \nbTripletMax{i} machines starting with a job of $\org_{4+5(i-1)}$ after the jobs from $\org_{2+5(i-1)}$ and $\org_{1+5(i-1)}$ such that each machine has a load of \UBOrgTwo{i}. Note that, as seen in \cref{claim:local_gap}, each job in \setIntegerTwo{i} scheduled this time interval has a processing time which is a multiple of \sumTripletInd{i} the same holds for the block of jobs in \setIntegerOne{i}. Finally the value $\UBOrgTwo{i}-\lengthOffset{i}$ is also a multiple of \sumTripletInd{i}, therefore that we can fill up completely the time between with jobs of processing time \sumTripletInd{i}. All jobs from $\org_{3+5(i-1)}$ complete at the latest at time \UBOrgTwo{i}, which is the optimal local makespan for $\org_{3+5(i-1)}$, i.e., the individual rationality is not violated.
\end{compactitem}

This completes the hardness proof.

}

By reducing from a DP-hard problem, we can show that the problem is beyond \NP\ and \coNP, even in the case where there are only two organizations. We conjecture that the problem remains $\thetaTwoPComplete$ in this case. %

\begin{restatable}[\appsymb]{proposition}{propMakespanDPk}
	\makespanMOSP\ is \DPhard even if $\nbOrg=2$.
	\label{prop:makespan_DPh_k}
\end{restatable}

\appendixproofwithstatement{prop:makespan_DPh_k}{\propMakespanDPk*}{

We reduce from the problem \probname{\ThreePartition AND NO-\ThreePartition} which we define now. This problem can be shown to be \DP-hard by a reduction from the \DP-complete problem \probname{SAT-UNSAT} \cite{papadimitriou1982complexity}, using standard reduction from \probname{SAT} to \ThreePartition \cite{garey1979computers}.

\decprob{\ThreePartition AND NO-\ThreePartition}{2 instances $\instTP=\{\{x_1,\dots,x_{3\nbTriplet}\}, \sumTriplet\}$ and $\instTPPrime=\{\{x'_1,\dots,x'_{3\nbTriplet'}\}, \sumTriplet'\}$ of \ThreePartition}{Is \instTP a yes-instance of \ThreePartition and \instTPPrime a no-instance of \ThreePartition?}

We assume both \sumTriplet and $\sumTriplet'$ to be even, if not, we multiply the values of the integers and the target sum by 2 in any instance in which the target sum is not even. We also assume that integers in the instances are strictly larger than the target sum divided by four and strictly smaller than the target sum divided by 2. We call $f=3\sumTriplet'/2+1$. We create an instance with 2 organizations:
\begin{compactitem}
	\item $\org_1$ owns \nbTriplet machines, for all $i \in [3\nbTriplet]$, it owns one job of processing time $x_i \cdot f$, it also owns one job of processing time $\sumTriplet'/2+1$.
	\item $\org_2$ owns $\nbTriplet'$ machines, for all $i \in [3\nbTriplet']$, it owns one ``short" job of processing time $x'_i$, it also owns $\nbTriplet'$ ``long" jobs of processing time $f\sumTriplet-3/2\sumTriplet'$.
\end{compactitem}
We set $\targetValue=f\sumTriplet$.

We start by observing that in a schedule with makespan \targetValue, each job of processing time $f\sumTriplet-3/2\sumTriplet'$ is on a distinct machine, indeed, if two such jobs were to be scheduled on the same machine, the total load of the machine would be of at least $f\sumTriplet-3\sumTriplet'+(3\sumTriplet'/2+1)\sumTriplet$, as $\sumTriplet$ is at least $3$, this would go beyond \targetValue.
Additionally, no machine can process both one job of processing time $f\sumTriplet-3/2\sumTriplet'$ and a job from $\org_1$, except the job of processing time $\sumTriplet'/2+1$, as it would have a load of at least $f\sumTriplet-3\sumTriplet'/2+(3\sumTriplet'/2+1) \cdot x$ where $x$ is an integer, which would go beyond \targetValue.

We now show that the \probname{\ThreePartition AND NO-\ThreePartition} instance is a yes-instance if and only if the reduced instance is a yes instance.

\onlyifDirection Let us first assume that the \probname{\ThreePartition AND NO-\ThreePartition} instance is a yes-instance. This means that \instTP is a yes instance, therefore, there exist a partition of its integers into triplets of sum \sumTriplet. This implies that there is a way of splitting jobs of $\org_1$, except the job of processing time $\sumTriplet'/2+1$, into triplets such that the total processing time of each triplet is precisely $f\sumTriplet$. We start building a schedule by scheduling each of these triplet on a distinct machine, note that since the total load of $\org_1$ is strictly larger than $f\sumTriplet\nbTriplet$ and since it owns $\nbTriplet machines$, its optimal local makespan is necessarily strictly greater than $f\sumTriplet$. 
Since \instTPPrime is a no-instance, it is impossible to partition its integers into triplets of sum \sumTriplet'. This implies that it is impossible to split short jobs of $\org_2$ into triplets of total processing time \sumTriplet'. This means that its optimal local makespan is strictly larger than $f\sumTriplet-3/2\sumTriplet'+\sumTriplet'$. Let us consider one optimal local schedule of $\org_2$, each machine schedules a long job and a set of short job. Since it is impossible to partition the short job perfectly at least one of the machines processes short jobs for at most $\sumTriplet'-1$. We reproduce this local schedule on \nbTriplet' machines to build~\sched, note that this does not violate individual rationality. Finally, we put the job of processing time $\sumTriplet'/2+1$ of $\org_1$ on a machine processing jobs of $\org_2$ for at most $f\sumTriplet-3/2\sumTriplet'+\sumTriplet'-1$, as seen previously, this machine is guaranteed to exist. The load of this machine is at most $f\sumTriplet-3/2\sumTriplet'+\sumTriplet'-1+\sumTriplet'/2+1=f\sumTriplet=\targetValue$. This means that all jobs from $\org_1$ are completed by time $f\sumTriplet$ which fulfills individual rationality.

\ifDirection We now assume that the reduced instance is a yes-instance. This means that there exists an individually rational schedule~\sched with makespan \targetValue. As seen earlier, this schedule cannot have on the same machine either two long jobs from $\org_2$ or one long job from $\org_2$ and one job of $\org_1$, except the job of processing time $\sumTriplet'/2+1$. This means that the $3\nbTriplet$ jobs of $\org_1$ are scheduled on the same $\nbTriplet$ machines. Since the makespan of the schedule is $f\sumTriplet$ and the sum of their processing time is $f\sumTriplet\nbTriplet$, it means that each machine is assigned a set of jobs of total processing time exactly $f\sumTriplet$, by construction, these sets are necessarily triplets, which implies that there exists a partition of the integers of \instTP into triplets of sum \sumTriplet, and therefore that \instTP is a yes-instance.
We now look at $\org_2$ and we assume towards a contradiction that \instTPPrime is a yes-instance. Since \instTPPrime is a yes-instance it is possible to partition the integers of \instTPPrime into triplets of sum $\sumTriplet'$, therefore it is possible to partition the short jobs of $\org_2$ into triplets of total processing time $\sumTriplet'$. This means that it is possible to build a local schedule in which each machine processes one long job and one triplet of jobs and have a load of $f\sumTriplet-\sumTriplet'/2$, this sets the optimal local makespan of $\org_2$ to $f\sumTriplet-\sumTriplet'/2$. Since~\sched is individually rational, no job of $\org_2$ can be completed in~\sched after $f\sumTriplet-\sumTriplet'/2$. This means that the job of processing time $\sumTriplet'/2+1$ has to be scheduled either after $f\sumTriplet-\sumTriplet'/2$ or after $f\sumTriplet$ on a machine with jobs from $\org_1$. In both cases this job would be completed after $f\sumTriplet$, a contradiction. This means that $\org_2$ is necessarily a no-instance.

This completes the proof.%
}

The next result shows that the problem remains \NPhard\ even for the case when finding an optimal local schedule for each organization is easy. The hardness persists even if each organization has only two jobs. The hardness proof follows from~\cite{cohen2011multi}.

\begin{restatable}[\appsymb]{proposition}{propMakespanNPmnmax}
	For constant $\maxNbTask$ or constant $\maxNbMachine$, \makespanMOSPDec\ is \NP-complete. It remains \NP-hard even if $\maxNbTask=2$ and $\maxNbMachine=1$.
	\label{prop:makespan_mmax_nmax}
\end{restatable}

\appendixproofwithstatement{prop:makespan_mmax_nmax}{\propMakespanNPmnmax*}{

  We start by observing that if $\maxNbTask$ is a constant or $\maxNbMachine$ is a constant, an optimal local-schedule can be computed in polynomial time. Let us first describe an algorithm when $\maxNbTask$ is a constant. If an organization has more than $\maxNbTask$ machines, then an optimal local schedule can be obtained by scheduling one job on each machine. Otherwise, one can brute-force search an optimal local schedule by checking all possible schedules of the $\maxNbTask$ jobs to at most $\maxNbTask$ machines, which is upper-bounded by $\maxNbTask^{\maxNbTask}$.

  Similarly, if $\maxNbMachine$ is a constant, one can check all the up to $\maxNbTask^{\maxNbMachine}$ possible local assignments of jobs to machines for each organization. If $\maxNbMachine$ is a constant, then this is polynomial. We can then compute an optimal local schedule in polynomial time and therefore check if a given schedule is individually rational in polynomial time. This concludes the containment proof.

  For the hardness proof, as already mentioned, it follows by using the same reduction of an NP-hardness result of \citeauthor{cohen2011multi}~\shortcite{cohen2011multi}.
  The reduction yields an instance where each organization has one machine and at most two jobs.
  Since in this case the optimal local-makespan of each organization is uniquely defined and can be determined directly and a schedule satisfying the local constraint of theirs is also individually rational, the correctness follows. 
}

\subsection{Algorithmic Results}
We start with a fairly straightforward FPT result for the number~$n$ of jobs.

\begin{restatable}[\appsymb]{proposition}{propMakespanFPTn}
	\makespanMOSP\ is \FPT\ \wrt $n$.
	\label{prop:makespan_FPT_n}
\end{restatable}

\appendixproofwithstatement{prop:makespan_FPT_n}{\propMakespanFPTn*}{

As we are not aware of any paper that showed the result for the standard setting, we start by showing that the local optimization problem can be solved in \FPT\ time \wrt $n$. Let $n'=|\setTaskOrg{i}|$ and $m'=|\setMachineOrg{i}|$ for organization $i$. We can distinguish two cases:
\begin{compactenum}
\item $\nbMachine'\geq n'$
\item $\nbMachine'< n'$
\end{compactenum}
In the first case, an optimal schedule assigns at most one job to each machine and can therefore be found in polynomial time. In the second case, one can upper bound the number of machines by $n'$ and therefore there are at most $(n'!)\cdot {n'}^{n'}$ possible schedules. Since the makespan for each of those schedules can be computed in polynomial time, it follows that the local problems are solvable in polynomial time.

We can therefore compute for each organization, the time by which all their jobs have to be done. This implicitly gives each job a deadline. For the global problem we can now use the same approach as for the local problem. We once again distinguish two cases:
\begin{compactenum}
\item $\nbMachine\geq n$
\item $\nbMachine< n$ 
\end{compactenum}
In the first case, an optimal schedule assigns each machine at most one job. This is obviously individually rational, as no organization can have a local makespan that is shorter than the longest processing time of a job that organization owns. This schedule can also be found in polynomial time, by assigning one job to each machine until no jobs are left.

In the second case, there are $(n!)\cdot {n}^{n}$ possible schedules, as the number of machines is upper-bounded by $n$. The individual rationality can be checked in polynomial time, as for each job the deadline was computed in the preprocessing step. Therefore, since the makespan of each schedule can also be computed in polynomial time it follows that the problem is solvable in time $(n!)\cdot {n}^{n}\cdot |\instTP|^{O(1)}$ and therefore \FPT\ \wrt $n$. 
}

Now, we turn to our main result: \cref{thm:pmax+k}.
As mentioned in the introduction, we extend the idea of the FPT algorithm by \citeauthor{Mnich15}~\shortcite{Mnich15}.
The idea is to group machines that for each processing time have the same number of jobs of that time together since jobs of the same processing time are interchangeable.
They observed that there is always a \myemph{balanced} optimal schedule.
Here, \myemph{balanced} means that all jobs of the same processing time can be evenly assigned among the machines so the difference is upper-bounded by a function in~$\maxProcTime$.
Due to this, the number of different groups is bounded by a function in~$\maxProcTime$.
Finally, one can design an ILP formulation that has an integer variable for each group specifying how many machines of that group exist in a balanced optimal schedule.

For the \MOSP\ setting, jobs belong to different organizations and may not be interchangeable, even if they have the same processing time.
We circumvent this by also considering the parameter ``the number~$\nbOrg$ of organizations''.
By grouping the jobs according to the \optimallocalmakespan\ of their organization and showing that for each group and each processing time, each machine has the same number of jobs of that processing time up to a difference of a function of $\maxProcTime$, we are able to design an ILP similarly to Mnich and Wiese.

Before we show \cref{thm:pmax+k},
we need two auxiliary lemmas and an observation and some additional definitions.
In the \makespanMOSP, each organization cares only about when its last job is finished.
This time cannot exceed their \optimallocalmakespan.
We order the jobs based on the \optimallocalmakespan\ of their organization.
We say two jobs~\taskOrg{i}{j} and \taskOrg{i'}{j'} belong to the same \myemph{phase} if their organizations have the same \optimallocalmakespan, i.e., $\localMakespan{i}=\localMakespan{i'}$. The jobs that belong to organizations with the smallest \optimallocalmakespan\ belong to phase $1$. Formally, phase $1$ consists of the jobs
$\{\taskOrg{i}{j}\mid \nexists i' \localMakespan{i'}<\localMakespan{i}\}$=$\bigcup_{\argmin\localMakespan{i}} J_i$. Similarly, the jobs that have the next smallest \optimallocalmakespan\ will be referred to as jobs in phase $2$ and so on. As all the jobs belonging to a single organization belong to the same phase it follows that the number of phases is upper-bounded by $\nbOrg$. Let $\phase(\taskOrg{i}{j})$ be the phase that job $\taskOrg{i}{j}$ belongs to. %
We define the \myemph{end of phase $\phaseVar$ for machine $z$ and schedule $\sched$} to be $\ndP_\sched(\phaseVar,z)=\max\{0\}\cup\{\completionTimeTaskSchedule{i}{j}{\sched} \mid \phase(\taskOrg{i}{j})\leq \phaseVar\wedge\machineTaskSchedule{i}{j}{\sched}=z\}$. Note that $0$ is added to the set, as it would be possible for the set to be empty otherwise.

We start with a simple observation that jobs in an individually rational schedule can be \myemph{well ordered}.
\begin{obs}\label{obs:phaseorder}
  For each individually rational schedule~$\sched$,
  there exists another individually rational schedule~$\sched'$ with makespan at most~$\makespanObj(\sched)$
  such that for each two jobs~$\alpha$ and $\beta$ that are assigned to the same machine,
  if $\alpha$ is in a phase earlier than $\beta$, then
  $\alpha$ is scheduled earlier than~$\beta$ as well.
\end{obs}
\begin{proof}
  Such a schedule~$\sched'$ can be found by iteratively switching
  consecutive jobs if they violate the well-ordered property.
  Each such exchange maintains individual rationality, as a job from a later phase
  belongs to an organization with larger optimal local-makespan and no job except the two which were exchanged in the ordering has a different completion time after this exchange.
  Repeating this process exhaustively yields the desired schedule~$\sched'$.
\end{proof}

By \cref{obs:phaseorder}, we assume from now on that every individually rational schedule satisfies the well-ordered property.
We utilize this to upper-bound the difference between completion times of each phase between two machines.
\begin{restatable}[\appsymb]{lemma}{lemPhaseDif}
  \label{lem:phasediff}
  For each individually rational schedule~$\sched$, %
  there exists an individually rational schedule~$\sched'$ with $\makespanObj(\sched')\leq \makespanObj(\sched)$ such that for each pair of machines $z_1$ and $z_2$ and for each phase~$\phaseVar$ it holds that $|\ndP_{\sched'}(\phaseVar,z_1)-\ndP_{\sched'}(\phaseVar,z_2)|\leq \maxProcTime^3+\maxProcTime$. 
\end{restatable}

\appendixproofwithstatement{lem:phasediff}{\lemPhaseDif*}{

   \renewcommand{\qedsymbol}{(of \cref{lem:phasediff})~$\diamond$}
  Let $z_1$ and $z_2$ be two machines and $\sched$ a schedule. 
Without loss of generality, let $\ndP_\sched(\phaseVar,z_1)-\ndP_\sched(\phaseVar,z_2)> \maxProcTime^3+\maxProcTime$. We will argue that in this case we can exchange jobs of phase $\phaseVar$ or lower in $z_1$ with jobs of a later phase in $z_2$, such that the difference $|\ndP_\sched(\phaseVar,z_1)-\ndP_\sched(\phaseVar,z_2)|$ shrinks. As this process can be repeated for as long as $\ndP_\sched(\phaseVar,z_1)-\ndP_\sched(\phaseVar,z_2)> \maxProcTime^3+\maxProcTime$ the lemma follows directly. Let $J_1$ be the set of jobs on machine $z_1$ that start at or after $\ndP_\sched(\phaseVar,z_2)$ and belong to phase $1$ to $\phaseVar$. We note that the sum of the  processing times of jobs in $J_1$ is at least $\maxProcTime^3$, as the jobs are scheduled without pause on a single machine and the first job starts at $\ndP_\sched(\phaseVar,z_2)+\maxProcTime$ at the latest. Let $J_2$ be the set of jobs on machine $z_2$ that start at $\ndP_\sched(\phaseVar,z_2)$ at the earliest and are completed at $\ndP_\sched(\phaseVar,z_1)$ at the latest. For $J_2$ there are two options:
  \begin{compactenum}
  \item The total length of jobs in $J_2$ is less than $\maxProcTime^3$
  \item The total length of jobs in $J_2$ is at least $\maxProcTime^3$.
  \end{compactenum}
  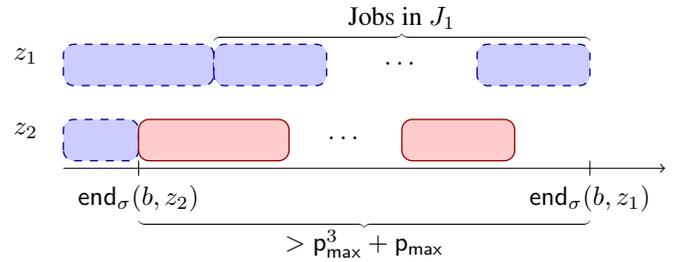
\begin{figure}[h]
  \begin{tikzpicture}
  	\draw [decorate, decoration = {calligraphic brace},rotate=180] (-2,-1.7) --  (-7,-1.7);
	\node at (4.5,2) {Jobs in $J_1$};
  	
	\begin{scope}[yshift=1cm]
		\taskColorTikz{ }{2}{2}{1}{colorOne};
		\taskColorTikz{ }{1.5}{3.5}{1}{colorOne};
		\node at (4.5,0.4) {$\dots$};
		\taskColorTikz{ }{1.5}{7}{1}{colorOne};  	
  	\end{scope}
  	
	\begin{scope}[yshift=0.0]
		\taskColorTikz{ }{1}{1}{1}{colorOne};  
		\taskColorTikz{ }{2}{3}{1}{colorTwo};  
		\node at (3.75,0.4) {$\dots$};
		\taskColorTikz{ }{1.5}{6}{1}{colorTwo};  	
  	\end{scope}
  	\draw[->] (0,0) -- (8,0);
  	\draw (7,0.1) -- (7,-0.1) node[below]{$\ndP_\sched(\phaseVar,z_1)$};
  	\draw (1,0.1) -- (1,-0.1) node[below]{$\ndP_\sched(\phaseVar,z_2)$};
  	\node at (-0.5,0.5) {$z_2$};
  	\node at (-0.5,1.5) {$z_1$};
  	\draw [decorate, decoration = {calligraphic brace,mirror},rotate=180] (-1,0.7) --  (-7,0.7);
	\node at (4,-1) {$>\maxProcTime^3+\maxProcTime$};
  \end{tikzpicture}
  \caption{Representation of a difference in end of phase $\phaseVar$ between two machines $z_1$ and $z_2$. Jobs of phases $\phaseVar$ and before are represented in blue with dashed lines, jobs from $J_2$ are represented in red with solid lines}
  \end{figure}
  In the first case, it follows that the difference in the makespan between the two machines is at least $\maxProcTime$. By picking an arbitrary job from $J_1$ making it start at $\ndP_\sched(\phaseVar,z_2)$ and shifting the other jobs on machine $z_2$ back, we therefore do not increase the total maximum makespan. Furthermore, this new schedule is also individually rational, as all jobs that were shifted back belong to a phase which is later than $\phaseVar$ and all jobs end at $\ndP_\sched(\phaseVar,z_1)$ at the latest. Therefore we have shown the statement in this case. 
  
  In the second case, as the total length of jobs in $J_1$ and $J_2$ is at least $\maxProcTime^3$ each, it follows that there is some processing time $p_{i_1}$ in $J_1$ such that there are at least $\maxProcTime$ many jobs of that processing time in $J_1$ per pigeonhole principle. Similarly, it follows that there is some processing time $p_{i_2}$ in $J_2$ such that there are at least $\maxProcTime$ many jobs of that processing time in $J_2$. We can then exchange $p_{i_1}$ many jobs of processing time $p_{i_2}$ in $J_2$ with $p_{i_2}$ many jobs of processing time $p_{i_1}$ in $J_1$. If we place the jobs from $J_2$ after all jobs from phases $1$ to $\phaseVar$ in $J_1$ we do not violate individual rationality, as they all end before $\ndP_\sched(\phaseVar,j_1)$. The jobs from $J_1$ will be placed directly after $\ndP_\sched(\phaseVar,j_2)$. As all jobs in $J_1$ start after $\ndP_\sched(\phaseVar,j_2)$, this also does not violate individual rationality. Afterwards, we can once again order the jobs on each machine according to phase. This concludes the proof of this Lemma.
}  
The next lemma upper-bounds the number of machines of the same type and phase.
Specifically, we upper- and lower-bound the number of jobs of each processing time and phase that can be assigned to a machine. %
  
\begin{restatable}[\appsymb]{lemma}{lemPhaseJob}
	\label{lem:phasejobs}

  Given an instance $\inst$ of \makespanMOSP\, let $J_{t,\phaseVar}$ be the set of jobs of processing time $t$ in phase $\phaseVar$. Then $\inst$ admits an optimal individually rational solution in which for 
  every phase $\phaseVar$ and every distinct processing time $t$ it holds that the number of jobs in $J_{t,\phaseVar}$ scheduled on each machine is in the range $[\lfloor\frac{|J_{t,\phaseVar}|}{m}\rfloor-O(\maxProcTime^{\maxProcTime}),\lfloor\frac{|J_{t,\phaseVar}|}{m}\rfloor+O(\maxProcTime^{\maxProcTime})]$.
\end{restatable}

\appendixproofwithstatement{lem:phasejobs}{\lemPhaseJob*}{
  This proof is a fairly minor modification of the lemma by~\citeauthor{Mnich15}~\shortcite{Mnich15}. Due to \cref{lem:phasediff} and \cref{obs:phaseorder} we can assume that the machines order the jobs according to phase, and the ending times of phases differ by at most $\maxProcTime^3+\maxProcTime$.

 Let $J_{t,\phaseVar}^{z}$ be the set of jobs of processing time $t$ in phase $\phaseVar$ that is scheduled on machine $z$. We show this lemma by showing that if the difference between the number of jobs of processing time $t$ belonging to phase $\phaseVar$ exceeds $2f(\maxProcTime)$ between two machines, we can exchange jobs until the difference is smaller than $2f(\maxProcTime)$. To this end, we show that there always exists an optimal solution, where $||J_{\ell,\phaseVar}^{z}|-|J_{\ell,\phaseVar}^{z'}||\leq h(\ell)\cdot g(\maxProcTime)$, where $h(\ell)=1+\sum_{v=\ell+1}^{\maxProcTime} v\cdot h(v)=\frac{(\maxProcTime+1)!}{(\ell+1)!}$ and $g(\maxProcTime)=3\maxProcTime^3+2\maxProcTime$. We show that in case a pair of machines $z,z'$ exist that violate this inequality for some phase $\phaseVar$ and some processing time of job $\ell$ we can exchange jobs of processing time $\ell$ with jobs of a shorter processing time in that phase, in order to satisfy this inequality. By repeating this process for all pairs of machines and all processing time this then leads to the result we are aiming for. We show that this exchange is possible by showing that there exists a certain lower bound on the total length of shorter jobs. 

Assume that $\ell$ is the largest processing time such that $|J_{\ell,\phaseVar}^{z}|-|J_{\ell,\phaseVar}^{z'}|> h(\ell)\cdot g(\maxProcTime)$ for a phase $\phaseVar$ and a pair of machines $z$ and $z'$ for the sake of readability we will omit $\phaseVar$ in the subscript, as it is fixed throughout the following inequalities:
\allowdisplaybreaks
\begin{align*}
&\sum_{t=1}^{\ell-1}t\cdot|J_t^{z'}|=(\sum_{t=1}^{\maxProcTime}t\cdot|J_t^{z'}|)-\ell\cdot|J_\ell^{z'}|-(\sum_{t=\ell+1}^{\maxProcTime}t\cdot|J_t^{z'}|)\\
&\geq(-2(\maxProcTime^3+\maxProcTime)+\sum_{t=1}^{\maxProcTime}t\cdot|J_t^{z}|)-\ell\cdot|J_\ell^{z'}|\\&-(\sum_{t=\ell+1}^{\maxProcTime}t\cdot|J_t^{z'}|)\\
&\geq(-2(\maxProcTime^3+\maxProcTime)+\sum_{t=1}^{\maxProcTime}t\cdot|J_t^{z}|)-\ell\cdot|J_\ell^{z'}|\\&-(\sum_{t=\ell+1}^{\maxProcTime}t\cdot|J_t^{z}|+h(t)\cdot g(\maxProcTime))\\
&>-2(\maxProcTime^3+\maxProcTime)+\sum_{t=1}^{\ell}t\cdot|J_t^{z}|+\ell(h(\ell)\cdot g(\maxProcTime)-|J_\ell^{z}|)\\&-\sum_{t=\ell+1}^{\maxProcTime}t\cdot h(t)\cdot g(\maxProcTime)\\
&=-2(\maxProcTime^3+\maxProcTime)+\sum_{t=1}^{\ell-1}t\cdot|J_t^{z}|+\ell(h(\ell)\cdot g(\maxProcTime))\\&-\sum_{t=\ell+1}^{\maxProcTime}t\cdot h(t)\cdot g(\maxProcTime)\\
&\geq\maxProcTime^3
\end{align*}
Note that the first inequality stems from the fact that due to \cref{lem:phasediff} we can assume that the difference between end points of a phase is at most $\maxProcTime^3+\maxProcTime$ and as the end point of the previous phase can also be shifted by the same amount we get the factor of $2$. 

Per pigeonhole principle it follows that there is a processing time $p<\ell$ such that jobs of that processing time sum up to $\maxProcTime^2$. Therefore there have to be at least $\maxProcTime$ many of those jobs. This allows us to exchange jobs of processing time $\ell$ with jobs of processing time $p$ without changing the total length of jobs in that phase on the machines, as the smallest common multiple of $p$ and $\ell$ is at most $p\cdot\maxProcTime$.

We can compute $f(\maxProcTime)=h(1)\cdot g(\maxProcTime)$. From~\citeauthor{Mnich15}~\shortcite{Mnich15} it follows that this is $2^{O(\maxProcTime\log\maxProcTime)}=O(\maxProcTime^{\maxProcTime})$, as the only difference is that $g(\maxProcTime)$ is slightly larger than in their paper (less than a factor $3$).
 This concludes the proof. 
   }
   
The observation and lemmas allow us to search for an optimal solution with a very specific structure. This allows us to formulate an ILP with FPT running time and leads to the following theorem:

\begin{theorem}
	\label{thm:pmax+k}
\makespanMOSP\ is \FPT\ \wrt $\maxProcTime +\nbOrg$ and therefore $\maxProcTime+\nbMachine$.
\end{theorem}
\begin{proof}
First note that $\nbOrg\leq \nbMachine$, as each organization has at least one machine. Therefore, it suffices to show the result for $\maxProcTime +\nbOrg$.
This approach is based on the FPT algorithm with respect to $\maxProcTime$ that solves $(P||\makespanObj)$ described by Mnich and Wiese~\shortcite{Mnich15}. 
Intuitively, the proof works by running an ILP that fixes the schedule for each phase and linking the phases together afterwards. 

We start by computing the number of phases by computing the optimal local makespan for each organization. By Mnich and Wiese's result this is doable in FPT time for each of the organizations. So computing it for all organizations is also doable in \FPT time. We now ignore the organizations and group jobs by phases as previously described. Let $[B]$ be the set of phases and $t_\phaseVar$ the latest time by which jobs of phase $\phaseVar$ must be finished. Let $P_\phaseVar$ be the jobs in phase $\phaseVar\in[B]$ and $P_\phaseVar^\ell$ the jobs with processing time $\ell$ in phase $\phaseVar$. For each phase $\phaseVar\in[B]$ we compute $y_\phaseVar:=\frac{|P_\phaseVar|}{m}$, this is the average makespan among the machines for jobs of only phase $\phaseVar$. Note that these precomputation steps are also doable in polynomial time once the optimal local solutions for the organizations have been computed. Due to \cref{lem:phasediff}, we know that for every machine $z$, we can require that a machine must satisfy that $\ndP_\sched(\phaseVar,z)\in[\sum_{\ell=1}^{\phaseVar}y_\ell-(\maxProcTime^3+\maxProcTime),\min{\sum_{\ell=1}^{\phaseVar}y_\ell+(\maxProcTime^3+\maxProcTime),t_\phaseVar}]$, for an optimal schedule $S$. Note that the left-hand side of the interval does not need a minimum, as assigning the optimal local schedule for each machine is individually rational. 

Similarly we can see that the first job of the phase $\phaseVar$ must be scheduled in $[\sum_{\ell=1}^{\phaseVar-1}y_\ell-(\maxProcTime^3+\maxProcTime),\min{\sum_{\ell=1}^{\phaseVar-1}y_\ell+(\maxProcTime^3+\maxProcTime),t_{\phaseVar-1}}]$, as we can assume that the jobs are ordered according to their phase due to \cref{obs:phaseorder}. 

We can now describe the constraints and variables of the integer linear program (ILP) that solves this problem instance. Note that we do not distinguish between jobs that belong to the same phase and type, in the following. We start by describing the variables:
\begin{compactitem}[--]
\item For each phase $\phaseVar\in[B]$, each possible starting point $\startVar\in[\sum_{\ell=1}^{\phaseVar-1}y_\ell-(\maxProcTime^3+\maxProcTime),\min\{\sum_{\ell=1}^{\phaseVar-1}y_\ell+(\maxProcTime^3+\maxProcTime),t_{\phaseVar-1}\}]$ (if $\phaseVar=1$ we fix $start=0$), each possible ending point $\ndVar\in[\sum_{\ell=1}^{\phaseVar}y_\ell-(\maxProcTime^3+\maxProcTime),\min\{\sum_{\ell=1}^{\phaseVar}y_\ell+(\maxProcTime^3+\maxProcTime),t_\phaseVar\}]$, and each vector $M$ of length $\maxProcTime$ that satisfies that $M_t$ is at least $\lfloor\frac{|P_\phaseVar^p|}{m}\rfloor-f(\maxProcTime)$ and at most $\lfloor\frac{|P_\phaseVar^p|}{m}\rfloor+f(\maxProcTime)$ and $\sum_{t=1}^{\maxProcTime} M_t\cdot t=\ndVar-\startVar$, we create a variable $v_{\phaseVar,\startVar,\ndVar,M}$. Intuitively, the vector $M$ keeps track of how many jobs of processing time $t$ are scheduled on a machine through the entry $M_t$. Note that all parameters must be non-negative integers (including zero) and that $M$ may be the zero vector. These variables must all take integer values in the range $[0,\nbMachine]$. Informally, the value this variable takes is the number of machines that finished the previous phase(s) at time $\startVar$, has exactly the number of jobs of each processing time as in $M$ scheduled on them in phase $\phaseVar$, and finishes phase $\phaseVar$ exactly at time $\ndVar$. 
\item We can also introduce an auxiliary variable $e$ in order to solve the optimization problem. This is not necessary.
\end{compactitem} 
These variables turn out to be the only variables that are needed. In this proof, parameters will be called valid if they can form a variable as described above. We now describe the constraints that are needed. 
\begin{compactenum}[(1)]
\item We need a constraint that limits the number of machines that can be used in each phase. As the total number of machines is $\nbMachine$, this simply means that the variables need to sum up to $\nbMachine$ for each fixed $\phaseVar\in[B]$. 
\begin{align*}
\forall~\phaseVar\in[B]\colon \sum_{\forall~\text{valid } \startVar,\phantom{i}\ndVar,\phantom{i}M} v_{\phaseVar,\startVar,\ndVar,M}=\nbMachine
\end{align*}
\item We need a constraint that makes sure that the starting times and end times of machines match between phases, such that each machine is ensured to only run one job at a time and not have any time when it is not processing any job. For the following constraint, let $C=\maxProcTime^3+\maxProcTime$.
\begin{align*}
&\forall~\phaseVar\in[B]\setminus\{1\},\forall~\mathsf{time}\in\\ &\big[\sum_{\ell=1}^{\phaseVar-1}y_\ell-C,\min\{\sum_{\ell=1}^{\phaseVar}y_\ell+C,t_\phaseVar\}\big] \colon\\ &\sum_{\forall~\text{valid } \startVar,\phantom{i}M^1} v_{\phaseVar-1,\startVar,\mathsf{time},M^1}=\\&\sum_{\forall~\text{valid } \ndVar,\phantom{i}M^2} v_{\phaseVar,\mathsf{time},\ndVar,M^2}
\end{align*}
\item We need a constraint that ensures that in each phase all the jobs that are part of this phase are scheduled. For each processing time $t$ we add:
\begin{align*}
\forall~\phaseVar\in[B]\colon \sum_{\forall~\text{valid } \startVar,\ndVar,M} M_t\cdot v_{\phaseVar,\startVar,\ndVar,M}=|P_i^t|
\end{align*}
\item In order to find an optimal solution we need to link the auxiliary variable $e$ to the other variables.
\begin{align*}
\forall~v_{\phaseVar,\startVar,\ndVar,M}: \min\{v_{\phaseVar,\startVar,\ndVar,M}\cdot \ndVar, \ndVar\}\leq e 
\end{align*}
\end{compactenum}
In order to solve the optimization problem for the makespan we can then minimize $e$ in the ILP.

We now show correctness of the ILP, by arguing that each valid schedule $\sched$ that has the form as described in \cref{obs:phaseorder}, \cref{lem:phasediff}, and \cref{lem:phasejobs} is a valid solution for the ILP (ignoring the minimization over $e$) and showing that every solution to the ILP can be transformed to a valid schedule $\sched$.

Let $\sched$ be a valid schedule, for each phase $\phaseVar$, $\startVar$, $\ndVar$, and $M$ we set $v_{\phaseVar,\startVar,\ndVar,M}$ to be equal to the number of machines that schedule jobs according to the vector $M$ in that phase, such that $\ndP_\sched(\phaseVar-1,z)=\startVar$. This satisfies constraint $1$, as each phase obviously only uses $\nbMachine$ machines. Constraint $2$ is satisfied, as we set $\ndP_\sched(\phaseVar-1,z)=\startVar$, and constraint $3$ is satisfied as we have a valid schedule.

For the other direction, we assign jobs phase by phase. For the first phase, we $v_{1,0,\ndVar,M}$ many machines with exactly the job seen in $M$. Then in step $\phaseVar$ we choose $v_{\phaseVar,\startVar,\ndVar,M}$ many machines that satisfy that $\ndP_\sched(\phaseVar-1,z)=start$ and assign the jobs in $M$ to them. This is necessarily possible, due to constraint $2$. Note that the number of machines in this step is exactly $\nbMachine$ due to constraint $1$ and all jobs in this phase are scheduled due to constraint $3$.

Finally, as $e$ only tracks the largest $end$ among variables $v_{\phaseVar,\startVar,\ndVar,M}\neq0$, it returns the makespan.

As the number of variables as well as the number of constraints is \FPT \wrt $\maxProcTime+\nbOrg$, it follows that the ILP solves the problem in \FPT time \wrt $\maxProcTime+\nbOrg$.
This concludes the proof. 
\end{proof}

The next two corollaries follow directly from the proof of \cref{thm:pmax+k},
as the number of phases and $\maxProcTime$ can be upper-bounded by the given parameters.

\begin{restatable}[\appsymb]{corollary}{corMakespanFPTtarget}
	\makespanMOSP\ is \FPT\ \wrt $\targetValue$.
	\label{cor:makespan_FPT_target}
\end{restatable}

\appendixproofwithstatement{cor:makespan_FPT_target}{\corMakespanFPTtarget*}{

It can be easily seen that $\targetValue$ upper-bounds $\maxProcTime$. However, the number of phases is also upper-bounded by $\targetValue$, as all organizations with local makespans that exceed $\targetValue$ can be grouped together. This holds because for them individual rationality is always upheld by a schedule with makespan $\targetValue$. 
}

\begin{restatable}[\appsymb]{corollary}{corMakespanFPTnmaxpmax}
	\makespanMOSP\ is \FPT\ \wrt $\maxNbTask+\maxProcTime$.
\label{cor:makespan_FPT_nmax_pmax}
\end{restatable}

\appendixproofwithstatement{cor:makespan_FPT_nmax_pmax}{\corMakespanFPTnmaxpmax*}{
  
As we assume that each organization has at least one machine it follows that each local makespan is upper-bounded by $\maxNbTask\cdot\maxProcTime$. As combining all local schedules leads to an individually rational schedule, it follows that if $\targetValue\geq\maxNbTask\cdot\maxProcTime$, there always exists an individually rational schedule. In the other case, $\targetValue$ can be upper-bounded by $(\maxNbTask+\maxProcTime)^2$ and this reduces to the previous corollary.
}

Finally, we use a dynamic programming approach that keeps track of the makespans of each machines for the assigned jobs in order to show the following result:
\begin{proposition}
	\makespanMOSP\ is \XP \wrt $\nbMachine$.
	\label{prop:makespan_XP_m}
\end{proposition}

\begin{proof}
  We first show how to compute an optimal local schedule for an arbitrary organization in \XP-time \wrt $\nbMachine$; we call this problem~\myemph{\minMakespan} via DP.
  Then, we show how to modify it to solve our problem~\makespanMOSP.
  For \minMakespan, we describe the DP for a given organization~$\org_i$ with jobs \setTaskOrg{i} and \setMachineOrg{i} machines.
  We note that the ordering of jobs on a machine does not matter, but rather their processing times.
  We go through the jobs of~$\org_i$ in this order~$\taskOrg{i}{1},\ldots,\taskOrg{i}{|\setTaskOrg{i}|}$. 

  We maintain a dynamic table~$\mathcal{D}_\loc(z_1,\ldots,z_{\nbMachineOrg{i}},j)\in\{0,1\}$, where $z_1,\ldots,z_{\nbMachineOrg{i}}\in[\sum_{\ell=1}^{|\setTaskOrg{i}|}\procTimeTask{i}{\ell} ]$ and $j\in[|\setTaskOrg{i}|]\cup\{0\}$.
  Intuitively, the table entry is $1$ if it is possible to assign the first~$j$ jobs
  to the machines such that machine~$d$, $d\in [\nbMachineOrg{i}]$ has makespan~$z_d$. We initialize the table with $\mathcal{D}_\loc(0,\ldots,0,0)=1$
We now describe the recurrence:
\begin{align*}
&\mathcal{D}_\loc(z_1,\dots,z_{\nbMachineOrg{i}},j)=\begin{cases}
  1, &\textbf{ if }\exists d\in [\nbMachineOrg{i}]\colon \mathcal{D}_\loc(z_1, \dots,\\
  & ~~z_d-\procTimeTask{i}{j},\dots,z_{\nbMachineOrg{i}},j-1)=1\\
  0, &\textbf{ else}
\end{cases}
\end{align*}
The correctness of the recurrence is straightforward since it branches over the options of where to assign the $j^{\text{th}}$ job.
Since the table has $(\nbTask\cdot \maxProcTime)^{m+1}$ entries, by finding the table entry $\mathcal{D}_\loc(z_1,\ldots,z_{m'},|\setTaskOrg{i}|)=1$ that minimizes $\max\{z_1,\ldots,z_{m'}\}$ an optimal schedule can be found. Therefore, we can solve \minMakespan\ and compute \optLocalMakespan{i} for each organization $\org_i$ in \XP-time \wrt $\nbMachine$. Note that we require \optLocalMakespan{i} for the individual rationality constraint. 

Now, we turn to our problem. Similarly to the proof of \cref{thm:pmax+k} we divide the jobs according to phases. As a reminder, $\phase(\taskOrg{i}{j})$ refers to the phase of $\taskOrg{i}{j}$. We order the jobs in a way $\task_1,\ldots,\task_n$, such that the jobs satisfy $\phase(\task_1)\leq\ldots\leq\phase(\task_n)$. Note that jobs from the same phase can be ordered in an arbitrary manner. We go through the jobs in this order.

We use $t(\phaseVar)$ to refer to the time by which jobs in phase $\phaseVar$ need to be done, i.e., the \optimallocalmakespan\ of the organizations whose jobs belong to this phase.

We maintain a dynamic table~$\mathcal{D}(z_1,\ldots,z_{\nbMachine},j)\in\{0,1\}$, where $z_1,\ldots,z_{\nbMachine}\in[\sum_{\ell=1}^{\sum_{i\in[k]}n_i}\procTimeTask{i}{\ell} ]$ and $j\in[\sum_{i\in[k]}n_i]\cup\{0\}$. Intuitively, the table entry is $1$ if it is possible to assign the jobs $\alpha_1,\ldots,\alpha_j$ to the machines such that machine $d$, $d\in[\nbMachine]$ has a makespan of $z_d$ and individual rationality is upheld.

We initialize the table with $\mathcal{D}(0,\ldots,0,0)=1$. We now describe the recursive step: 
\begin{align*}
	&\mathcal{D}(z_1,\ldots,z_{\nbMachine},j)=\\&\begin{cases}
		1, &\textbf{ if }\exists d\in [\nbMachine]\colon \mathcal{D}(z_1, \dots,\\
		& ~~z_d-\procTimeTask{}{j},\dots,z_{\nbMachine},j-1)=1\\
		&\textbf{and } \max\{z_1,\ldots,z_{\nbMachine}\}\leq t(\phase(\alpha_j))\\
		0&\textbf{, else}
	\end{cases}
\end{align*}

The correctness of the recurrence is straightforward since it branches over the options of where to assign the $j^{\text{th}}$ job. We can do this, as we can assume that the optimal schedule is well-ordered due to \cref{obs:phaseorder} and the ordering of jobs in the same phase on the same machine does not matter for a well-ordered schedule. Individual rationality is guaranteed, as the job must be done before the deadline due to $\max\{z_1,\ldots,z_{\nbMachine}\}\leq t(\phase(\alpha_j))$. Since the table has $(\nbTask\cdot \maxProcTime)^{m+1}$ entries, by finding the table entry $\mathcal{D}_(z_1,\ldots,z_{m},n)=1$ that minimizes $\max\{z_1,\ldots,z_{m}\}$ an optimal schedule can be found. Therefore \makespanMOSP\ is in \XP\ \wrt $\nbMachine$.
\end{proof}

\section{Minimizing the Sum of Completion Times}
\label{sec:sumC}
\appendixsection{sec:sumC}

While \makespanMOSP\ inherits hardness from the matching scheduling problem, minimizing the makespan, it is not clear that \sumCMOSP\ is \NP-hard. Indeed, in the traditional scheduling setting, a schedule with minimum sum of completion times can be found in polynomial time~\cite{brucker1999scheduling}. We show that \sumCMOSPDec\ is \NP-complete, even for a constant maximum number of jobs (resp. machines) per organization.

\begin{restatable}[\appsymb]{theorem}{thmSumCNP}
	\sumCMOSPDec\ is \NPcomplete. It remains \NP-hard even if $\maxNbTask=3$ and $\maxNbMachine=2$.
	\label{thm:avg_NP}
\end{restatable}

\begin{proof}[Proof sketch]
Containment follows from the fact that optimal local schedules can be computed in polynomial time. For hardness, we reduce from the \NP-complete problem \ThreePartition which aims at partitionning a set of integers into triplets of the same sum \sumTriplet. We will create ``triplet" organizations which benefit from the cooperation by starting one of their jobs earlier. A triplet organization can then accept to delay its jobs but only by a total processing time \sumTriplet, otherwise the schedule would not be individually rational. Other organizations own ``integer" jobs with processing times matching the integers from the \ThreePartition instance. To meet the sum of completion times objective, it will be necessary to schedule an integer job first on all machines, therefore delaying jobs from triplet organizations. To both fulfill individual rationality and meet the sum of completion times objective, it will be necessary to delay jobs from each triplet organizations by exactly \sumTriplet, which is only possible if the integers can be partitioned into triplets of sum \sumTriplet. 
\end{proof}

\appendixproofwithstatement{thm:avg_NP}{\thmSumCNP*}{

\setcounter{theorem}{3}

We start by arguing that the problem is contained in NP. We can easily check in polynomial time that a given solution is individually rational, as the schedules are given, and that the total sum of completion times is indeed lower than or equal to the target.

To prove hardness, we reduce from the \ThreePartition problem that we introduce now.

\decprob{\ThreePartition}{An integer \sumTriplet, a set \setInt of 3\nbTriplet integers $\{\integer_1,\dots,\integer_{3\nbTriplet}\}$ such that $\sum\limits_{i \in [3\nbTriplet]} \integer_i=\nbTriplet\sumTriplet=\totsum$.}{Is there a partition of \setInt into \nbTriplet triplets $\{\triplet_1,\dots,\triplet_{\nbTriplet}\}$, such that the sum of the integers in each triplet is exactly \sumTriplet?}

The \ThreePartition problem is NP-hard even if all the integers in \setInt have value between $\sumTriplet/4$ and $\sumTriplet/2$; $\sumTriplet/4$ and $\sumTriplet/2$ not included \cite{garey1979computers}. We make such an assumption for the reduction.

In the reduced instance, we create:
\begin{compactitem}
	\item $\nbTriplet$ ``integer'' organizations labeled from $\org_1$ to $\org_{\nbTriplet}$. For all $i$ in $[\nbTriplet]$, organization $\org_i$ owns 1 machine and three jobs $\taskOrg{i}{1},\taskOrg{i}{2},\taskOrg{i}{3}$ and for all $j$ in $[3]$, $\procTimeTask{i}{j}=x_{3(i-1)+j}$.
	\item $\nbTriplet$ ``triplet'' organizations, labeled from $\org_{\nbTriplet+1}$ to $\org_{2\nbTriplet}$. Each triplet organization owns 2 machines and 3 jobs of processing time $\sumTriplet$.
\end{compactitem}
The local schedules are obtained by running the SPT (Shortest Processing Time) list scheduling algorithm, i.e., the jobs are sorted by non decreasing processing time and scheduled greedily~\cite{brucker1999scheduling}. 

The local sum of completion times of each triplet organization is precisely $4\sumTriplet$. The local sum of completion times of the integer organizations is larger than the sum of processing times of the jobs owned by the organization as the organization only owns 1 machine but three jobs.

We set $\targetValue=5\totsum$. This concludes the construction.

Intuitively, a schedule with a total sum of completion times of $5\totsum$ is necessarily a schedule in which all jobs owned by integer organization are scheduled first on a machine and the jobs of the triplet organization are all scheduled second on a machine. However, such a schedule is individually rational only if the sum of completion times of the jobs owned by triplet organizations is at most $4\sumTriplet$, as the three jobs of processing time $\sumTriplet$ of any triplet organization are delayed by three jobs owned by integer organizations, the sum of the completion times of these jobs have to be at most $\sumTriplet$. As the total sum of processing times of jobs owned by integer organizations is precisely $\nbTriplet\sumTriplet$, each group of three jobs delaying jobs from one of the $\nbTriplet$ triplet organizations need to have processing times summing to precisely $\sumTriplet$. 

We will now prove that the reduced instance is a yes-instance if and only if the \ThreePartition instance is a yes-instance.

\onlyifDirection We first prove a claim. Note that the solution described in this claim does not necessarily satisfy individual rationality. 

\begin{claim}
A schedule minimizing the total sum of completion times assigns exactly two jobs to each machine. The job starting first on a machine is owned by an integer organization, the second by a triplet organization.
\label{claim:sumC_structure_schedule}
\end{claim}

\begin{proof}\renewcommand{\qed}{\hfill (end of the proof of~\cref{claim:sumC_structure_schedule})~$\diamond$}
Let us assume for the sake of contradiction that a schedule $\sched^*$ minimizing the total sum of completion times does not assign 2 jobs to each machine. Since there are $6\nbTriplet$ jobs and $3\nbTriplet$ machines, $\sched^*$ assigns at least 3 jobs to a machine and at most 1 job to another machine. We consider a machine $m_1$ which is assigned at least three jobs in $\sched^*$, and a machine $m_2$ which is assigned at most one job in $\sched^*$. On machine $m_1$, we can assume that jobs are scheduled by non-decreasing processing time, as otherwise a simple exchange between two consecutive jobs would decrease the sum of completion times. Therefore, if we call $\task_1,\task_2$ and $\task_3$ the jobs scheduled respectively first, second and third on $m_1$, we can assume that $\procTimeTask{}{\task_1} \leq \procTimeTask{}{\task_2} \leq \procTimeTask{}{\task_3}$. 
We consider the schedule $\sched'$ similar to $\sched^*$ except that $\task_1$ is scheduled first on $m_2$, i.e., the job scheduled alone on $m_2$ in $\sched^*$ now starts after $\task_1$; and jobs $\task_2$ and $\task_3$ start $\procTimeTask{}{\task_1}$ earlier on $m_1$. We now argue that $\sumCSched{\sched'}<\sumCSched{\sched^*}$. Indeed, the completion times of $\task_2$ and $\task_3$ are lower in $\sched'$ than in $\sched^*$ by $\procTimeTask{}{\task_1}$ and the completion time of the job scheduled alone on $m_2$ in $\sched^*$ is increased by $\procTimeTask{}{\task_1}$ in $\sched'$. The completion time of all other jobs is the same in the two schedules. This implies that $\sumCSched{\sched'}<\sumCSched{\sched^*}$, a contradiction.

We conclude this proof by noting that each machine is assigned two jobs, one owned by an integer organization and one from a triplet organization. Indeed, let us assume that a schedule minimizing the sum of completion times $\sched^*$ assigns two jobs owned by integer organizations to the same machine $m_1$. By pigeonhole principle, it also assigns two jobs owned by triplet organizations to another machine $m_2$. Since all integers in \setInt have a value strictly lower to $\sumTriplet/2$ the completion time of the last job on $m_1$ is strictly lower than \sumTriplet, therefore moving the second job scheduled on $m_2$ to the last position in $m_1$ would strictly decrease the sum of completion times, a contradiction. Furthermore, the job owned by the integer organization is scheduled first as it has a processing time strictly lower than the job owned by the triplet organization.
\end{proof}

We now argue that the minimum sum of completion times is of precisely $5\totsum$. Indeed, the sum of completion times of all jobs owned by integer organizations is precisely $\totsum$, as each of these jobs is scheduled first on a machine. The sum of completion times of jobs owned by triplet organizations is of $3\totsum$, which correspond to the sum of their processing time, plus the delay caused by the jobs of other organizations, which sums to precisely $\totsum$; for a total of $4\totsum$ for jobs owned by triplet organizations and a total or $5\totsum$ for all jobs. This means that any schedule with a sum of completion times of $5\totsum$ necessarily schedules jobs of integer organizations first, one on each machine, and then jobs owned by triplet organizations afterwards, one per machine.

While such a schedule obviously exists, it is not clear that an individually rational schedule can fulfill this condition. We will now show that if such an individually rational schedule exists, then the instance of \ThreePartition is a yes-instance.

Let us assume that there exists a schedule~\sched such that~\sched is individually rational and $\sumCSched{\sched}=5\totsum$. Since $\sumCSched{\sched}=5\totsum$, in~\sched each machine is assigned a job owned by an integer organization, scheduled first, and another job owned by a triplet organization scheduled afterwards. We suppose without loss of generality that the jobs of organization $\org_{q+i}$ are scheduled on machines $m_{3(i-1)+1}$, $m_{3(i-1)+2}$ and $m_{3(i-1)+3}$. Since~\sched is individually rational, the sum of completion times of jobs of $\org_{q+i}$ is at most $4\sumTriplet$, this is only possible if the sum of the processing times of the jobs owned by integer organizations assigned to machines $m_{3(i-1)+1}$, $m_{3(i-1)+2}$ and $m_{3(i-1)+3}$ sum to at most $\sumTriplet$. As this applies to organization $\org_{q+i}$ for all $i$ in $[\nbTriplet]$ and the sum of processing times of all jobs of integer organizations sum to $\nbTriplet\sumTriplet$, it means that for all $i$ in $[\nbTriplet]$ the sum of processing times of the three jobs owned by integer organizations assigned to $m_{3(i-1)+1}$, $m_{3(i-1)+2}$ and $m_{3(i-1)+2}$ sum to precisely $\sumTriplet$. We call these three jobs $\taskOrg{l}{j}$, $\taskOrg{l'}{j'}$, and $\taskOrg{l''}{j''}$. By construction, we have that the triplet $T_i=\{x_{3(l-1)+j}+x_{3(l'-1)+j'}+x_{3(l''-1)+j''}\}$ sums to \sumTriplet. This gives us a valid three partition.

\ifDirection Let us assume that the instance of \ThreePartition is a yes-instance. We build the schedule~\sched as follows: For all $i$ in $[q]$, we consider the triplet $\triplet_i=\{x_{3(l-1)+j},x_{3(l'-1)+j'},x_{3(l''-1)+j''}\}$, with $j,j'$ and $j''$ in $[3]$ and put the jobs \taskOrg{l}{j},\taskOrg{l'}{j'}, and \taskOrg{l''}{j''} first on machines $m_{3(i-1)+1}$,$m_{3(i-1)+2}$, and $m_{3(i-1)+3}$ respectively, we schedule the jobs of organization $\org_{\nbTriplet+1}$ on $m_{3(i-1)+1}$,$m_{3(i-1)+2}$, and $m_{3(i-1)+3}$ respectively after the jobs already scheduled. We now argue that the schedule~\sched is individually rational. It is straightforward to see that no integer organization has a larger sum of completion times in~\sched than in its local schedules as all of its jobs are scheduled first on a machine. The sum of completion times of each triplet organization is precisely $4\sumTriplet$ by the same argument used earlier: the sum of processing times of the jobs scheduled before the jobs of any triplet organization is precisely \sumTriplet. This means that the three jobs of processing time \sumTriplet owned by any given triplet organization are delayed by \sumTriplet in~\sched, which gives a sum of completion times of $4\sumTriplet$ for the organization, which is precisely its local sum of completion times. Furthermore, by the previous observation about the sum of completion times, since~\sched assigns to each machine one job owned by an integer organization, scheduled first on the machine, and one job owned by a triplet organization scheduled afterwards, the sum of completion times of~\sched is $5\totsum$. Therefore the reduced instance is a yes-instance. 
}
Using a similar idea as for \cref{thm:avg_NP}, we show that \sumCMOSP\ is W[1]-hard \wrt\ \nbOrg, i.e., it is unlikely to be in \FPT\ according to current complexity assumptions.

\begin{restatable}[\appsymb]{proposition}{propSumWOnem}
\sumCMOSP\ is W[1]-hard \wrt\ \nbOrg.
\label{prop:avg_W1_m}
\end{restatable}

\appendixproofwithstatement{prop:avg_W1_m}{\propSumWOnem*}{

\setcounter{theorem}{5}

We reduce from the \UnaryBinPacking\ problem that we introduce now.

\decprob{\UnaryBinPacking}{A set \setInt of $y$ integers $\{\integer_1,\dots,\integer_{y}\}$, an integer \sumTriplet, an integer \nbBin.}{Is there a partition of \setInt into \nbBin sets $\{\setBin_1,\dots,\setBin_{\nbBin}\}$, such that the sum of the integers in each set is lower or equal to \sumTriplet?}

The \UnaryBinPacking\ problem is W[1]-hard \wrt\ the number of bins, i.e. \nbBin \cite{jansen2013bin}. We assume $y>\nbBin$ as otherwise, we can simply put one integer in each set. Furthermore, we assume that no integer has value strictly greater than \sumTriplet, as otherwise the answer is necessarily no, additionally, if an integer has value precisely \sumTriplet, we have to put it alone in a set and we can reduce the instance to the same with one set and the integer removed.

In the reduced instance, we create:
\begin{compactitem}
	\item 1 ``integer'' organization $\org_1$, it owns \nbBin machines. For all $i$ in $[y]$, organization $\org_1$ owns one job $\taskOrg{1}{i}$ of processing time $\procTimeTask{1}{i}=\integer_{i}$.
	\item $\nbBin$ ``bin'' organizations, labeled from $\org_{2}$ to $\org_{\nbBin+1}$. Each bin organization owns $y$ machines and $y+1$ jobs of processing time $\sumTriplet$.
\end{compactitem}
Optimal local schedules are obtained by running the SPT (Shortest Processing Time) list scheduling algorithm~\cite{brucker1999scheduling}. 

The local sum of completion times of each bin organization is precisely $(y+2)\sumTriplet$. The local sum of completion times of the integer organization is larger than the sum of processing times of the jobs owned by the organization as the organization only owns \nbBin machine but \nbTask jobs.

We set $\targetValue=\nbBin(\sumTriplet+1)+2\sum_{i \in [y]}\procTimeTask{1}{i}$. This concludes the construction. Note that in the reduced instance, we have $\nbOrg=f(\nbBin)=\nbBin+1$. 

Intuitively, a schedule with a total sum of completion times of $\targetValue$ is necessarily a schedule in which all jobs owned by the integer organization are scheduled first on a machine and the jobs of the bin organization are scheduled afterwards. However, such a schedule is individually rational only if the sum of completion times of the jobs owned by bin organizations is at most $(\nbTask+2)\sumTriplet$, which is only possible if the processing times of jobs scheduled before its jobs is no more than \sumTriplet, which would correspond to a feasible set of the \UnaryBinPacking\ problem.

We will now prove that the reduced instance is a yes-instance if and only if the \UnaryBinPacking\ instance is a yes-instance.

\onlyifDirection We first prove a claim.

\begin{claim}
A schedule with total sum of completion times $\nbBin(\sumTriplet+1)+2\sum_{i \in [y]}\procTimeTask{1}{i}$ schedules the jobs of the integer organization first on separate machines and jobs from bin organizations later.
\label{claim:sumCBinPacking_structure_schedule}
\end{claim}

\begin{proof}\renewcommand{\qed}{\hfill (end of the proof of~\cref{claim:sumCBinPacking_structure_schedule})~$\diamond$}
Let us assume for the sake of contradiction that a schedule~\sched with total sum of completion times $\nbBin(\sumTriplet+1)+2\sum_{i \in [y]}\procTimeTask{1}{i}$ does schedule the jobs of the integer organization first on separate machines and jobs from bin organizations afterwards. First note that the sum of processing times in the instance is $\nbBin(\sumTriplet+1)+\sum_{i \in [y]}\procTimeTask{1}{i}$, this means that if all tasks could start at time 0, the sum of completion times would be $\nbBin(\sumTriplet+1)+\sum_{i \in [y]}\procTimeTask{1}{i}$. Now, the instance has $\nbBin\cdot(y+1)+y$ jobs and $\nbBin\cdot(y+1)$ machines. This means that at least $y$ jobs are delayed by at least another job. By scheduling two jobs on the same machine, the second one is delayed by the processing time of the first one, therefore, by scheduling the jobs of the integer organization first and on disjoint machines, the delay on the jobs owned by bin organizations is in total of $\sum_{i \in [\nbTask]}\procTimeTask{1}{i}$. Note that it is not possible to have a total sum of completion times lower, as this schedule could be obtained by an SPT algorithm, which is optimal for the minimization of the sum of completion times.
\end{proof}

While such a schedule obviously exists, it is not clear that an individually rational schedule can fulfill this condition. We will now show that if such an individually rational schedule exists, then the instance of \UnaryBinPacking\ is a yes-instance.

Let us assume that there exists a schedule~\sched such that~\sched is individually rational and $\sumCSched{\sched}=\targetValue$. Since $\sumCSched{\sched}=\targetValue$, jobs of the integer organization are scheduled first and jobs owned by bin organizations are scheduled afterwards. Since~\sched is individually rational, the sum of completion times of jobs of any bin organization is at most $(y+2)\sumTriplet$, this is only possible if the sum of the processing times of the jobs owned by integer organizations assigned to machines processing one of the jobs of a bin organization sums to at most $\sumTriplet$, as the sum of processing times of the tasks of each bin organization is $(y+1)\sumTriplet$. As this applies to all bin organizations, it means that for all $i$ in $[\nbBin]$ the sum of processing times of the jobs owned by the integer organization assigned to machines processing tasks of $\org_{1+i}$ sum to at most $\sumTriplet$. We consider the set $\setBin_i$ of these jobs, by construction $\sum_{\taskOrg{1}{j} \in \setBin} x_{j} \leq \sumTriplet$. As all jobs owned by the integer organization are scheduled, and as there are \nbBin bin organizations, there are \nbBin sets of jobs, and therefore \nbBin sets of integers summing to at most \sumTriplet. This gives us a valid bin packing.

\ifDirection Let us assume that the instance of \UnaryBinPacking is a yes-instance. We build the schedule~\sched as follows: For all $i$ in $[\nbBin]$, we consider the set $\setBin_i$ and put the jobs $\{\taskOrg{1}{j}|x_j \in \setBin_i\}$ first on machines owned by $\org_i$ we schedule the jobs of organization $\org_i$ on the same machines afterwards, except for one job, which is scheduled first on a machine owned by the integer organization. We now argue that the schedule~\sched is individually rational. It is straightforward to see that the integer organization does not have a larger sum of completion times in~\sched than in its local schedule as all of its jobs are scheduled first on a machine. The sum of completion times of each bin organization is at most $\sumTriplet \cdot (y+2)$ by the same argument used earlier: the sum of processing times of the jobs scheduled before the jobs of any bin organization is at most \sumTriplet. This means that the jobs of processing time \sumTriplet owned by any given bin organization are delayed by \sumTriplet at most in~\sched, which gives a sum of completion times of $\sumTriplet \cdot (y+2)$ for the organization, which is precisely its local sum of completion times. Furthermore, by the previous observation about the sum of completion times, since~\sched schedules jobs owned by the integer organization first on disjoint machines and schedules jobs of bin organizations afterwards, the sum of completion times of~\sched is $\nbBin(\sumTriplet+1)+2\sum_{i \in [y]}\procTimeTask{1}{i}$, therefore the reduced instance is a yes-instance.

This completes the parameterized reduction.
}

Similarly to \cref{prop:makespan_FPT_n}, the sum of completion times case allows for an \FPT\ result \wrt $\nbTask$ using a simple brute-forcing approach.
\begin{restatable}[\appsymb]{proposition}{propSumFPTn}
\sumCMOSP\ is \FPT\ \wrt $\nbTask$.
\label{prop:avg_FPT_n}
\end{restatable}

\appendixproofwithstatement{prop:avg_FPT_n}{\propSumFPTn*}{

This proof works in the same way as the proof for \cref{prop:makespan_FPT_n}. We can distinguish two cases:\begin{compactenum}
\item $\nbMachine\geq n$
\item $\nbMachine<n$
\end{compactenum}
In the first case, an optimal schedule assigns at most one job to each machine again and can be found in linear time. In the second case, as $\nbMachine<n$ it follows that there are at most $n!\cdot n^n$ different possible schedules and since the sum of completion times can be computed in polynomial time for each of them and individual rationality can also be computed in polynomial time, the statement follows.
}

As the $\targetValue$ upper-bounds the number of jobs, the following corollary follows directly.

\begin{restatable}[\appsymb]{corollary}{propSumFPTtarget}
\sumCMOSP\ is \FPT\ \wrt $\targetValue$.
\label{prop:avg_FPT_target}
\end{restatable}

\appendixproofwithstatement{prop:avg_FPT_target}{\propSumFPTtarget*}{

It holds that $\targetValue>\nbTask$, as each job has processing time at least $1$ and each job has a completion time that is at least its processing time. Therefore this result follows directly from \cref{prop:avg_FPT_n}.
}

Finally, we use a dynamic programming approach similar to the one used in the proof of \cref{prop:makespan_XP_m}. As it is not possible to order the jobs according to phase, as it was done for the \makespanMOSP\ case, we require an additional parameter for the dynamic programming approach to function.

\begin{restatable}[\appsymb]{proposition}{propSumCXPpmaxm}
\label{prop:avg_XP_m}
\sumCMOSP\ is \XP\ \wrt $\maxProcTime+\nbMachine$.
\end{restatable}

\appendixproofwithstatement{prop:avg_XP_m}{\propSumCXPpmaxm*}{

This result can be shown by modifying the proof of \cref{prop:makespan_XP_m} slightly. Intuitively, the modification is necessary, as it is not possible to group the jobs in phases, as an organization does not mind if a job is finished late, as long as another job finishes earlier to cancel this out.

We note that $\nbMachine>\nbOrg$, as each organization owns at least one machine. For each organization $\org_i$, let $\setTaskOrg{i}^\ell$ denote the set of jobs of processing time $\ell$. 
 Intuitively, we will modify the DP from \cref{prop:makespan_XP_m} by keeping track of how many jobs from each set $\setTaskOrg{i}^\ell$ have been assigned and the sum of completion times for each machine and organization. We compute the function $\mathcal{D}(z_1,\ldots,z_{m},t_1,\ldots,t_k,j_1^1,\ldots,j_1^{\maxProcTime},\ldots,j_k^{\maxProcTime})\in\{0,1\}$, where $z_1,\ldots,z_{m}\in[\sum_{\ell=1}^{(|\setTaskOrg{i}|}\procTimeTask{i}{\ell})^2 ]$ and $j_i^\ell\in[|\setTaskOrg{i}^\ell|]\cup\{0\}$. Intuitively, this function will take the value $1$, if it is possible to assign $j_i^\ell$ jobs from $\setTaskOrg{i}^\ell$ for all $i,\ell$ to the machines such that machine $d$ has a makespan of $z_d$ and organization $i$ has a local sum of completion time of $t_i$ restricted to the currently assigned jobs. We initialize the table in the following way:
\begin{align*}
\mathcal{D}(0,\ldots,0)=1
\end{align*}
Intuitively, this can be read as if we assign no jobs then all machines and organizations have a sum of completion times of $0$.
We now describe the recursive step:
\begin{align*}
&\mathcal{D}(z_1,\ldots,z_{\nbMachine},t_1,\ldots,t_k,j_1^1,\ldots,j_1^{\maxProcTime},\ldots,j_k^{\maxProcTime})=\\
&\begin{cases}
1&\text{, if }\exists i\in[\nbOrg], d\in[m],\ell\in[\maxProcTime]\colon \\
&\mathcal{D}(\ldots,z_m-\ell,\ldots,t_i-z_m,\ldots,j_i^\ell-1,\ldots)=1\\
0&\text{, else}
\end{cases}
\end{align*}
An optimal solution is a can then be found by finding a table entry with value $1$, such that $j_i^\ell=|\setTaskOrg{i}^\ell|$ for all $i,\ell$ that satisfies that $t_i$ is smaller or equal than the local sum of completion times and minimizes $\sum_{i\in[k]}t_i$. 

We first show that each table entry satisfies that it is a partial solution, i.e., if only $j_i^\ell$ jobs in $\setTaskOrg{i}^\ell$ are assigned for each $i,\ell$, then there is a solution such that each machine $d$ has makespan $z_d$ and each organization $\org_i$ has local sum of completion times $t_i$. We show this inductively. The base case where every parameter is $0$ obviously holds. Then the correctness follows via the recursion, as it models a task of processing time $\ell$ belonging to organization $\org_i$ being assigned to machine $d$.

On the other hand when given a schedule one can find a corresponding table entry, by step by step assigning the jobs to each machine.

As the table has at most $(n\cdot\maxProcTime)^m+(n^2\cdot\maxProcTime)^m+n^{m\cdot\maxProcTime}$ entries and each entry can be computed in polynomial time, it follows that the problem is solvable in \XP-time \wrt $\maxProcTime+\nbMachine$.

This concludes the proof.
}

\section{Conclusion}\label{sec:conclusion}
We introduce the concept of individual rationality into multi-organizational scheduling and explore the parameterized complexity of two optimization problems, \makespanMOSP\ and \sumCMOSP.
For the former problem, an important open question remains: Is the problem fixed-parameter tractable (FPT) with respect to the maximum completion time~$\maxProcTime$? Notably, the classical single-organization variant of this problem is already known to be FPT with respect to~$\maxProcTime$.

Our research opens up several promising avenues for future work.
First, an immediate extension is to consider the case of parallel jobs.
Second, our framework can be applied to other scheduling problems, such as those with precedence constraints or hard deadlines.
Third, it would be also interesting to study scenarios where each organization provides a precomputed local schedule as input.
For  \sumCMOSP, the complexity remains unchanged, as optimal local schedules can be computed in polynomial time.
For  \makespanMOSP, this setting reduces the problem to within NP, and preliminary investigations suggest that the parameterized results carry over.

Finally, an intriguing direction is to study scenarios where the local and global objectives differ.
For instance, the global objective might be to minimize~$\sumCObj$, while individual rationality mandates that the makespan of each organization matches its locally optimal makespan. A related model was previously examined by~\citeauthor{cohen2011multi}~\shortcite{cohen2011multi}, but without considering individual rationality as a constraint.

\clearpage
\section*{Acknowledgements}
The authors are supported by the Vienna Science and Technology Fund (WWTF)~[10.47379/ VRG18012]. We would like to thank the reviewers for their helpful comments.
\bibliographystyle{named}
\bibliography{bibliography}

\clearpage
\begin{table}[t!]
  \centering
  \Large \textbf{\appendixtitle}
\end{table}
\bigskip

\begin{appendix}
\appendixtext
\end{appendix}

\end{document}